\documentclass[aps,showkeys,pra,reprint,superscriptaddress,groupaddress,floatfix,byrevtex]{revtex4-1}
\usepackage{graphicx}
\usepackage{dcolumn}
\usepackage{bm}
\usepackage[utf8x]{inputenc}
\usepackage{srctex}
\usepackage{amsmath}
\usepackage{amsfonts}
\usepackage{amssymb}
\usepackage{graphicx}
\usepackage{srctex}
\usepackage[dvipsnames]{xcolor}
\usepackage{subfigure}
\usepackage{cancel}
\usepackage{mathtools}

\newcommand{\beq}{\begin{equation}}
\newcommand{\eeq}{\end{equation}}
\definecolor{rosita}{rgb}{0.97, 0.56, 0.65}
\newcommand{\ket}[1]{\ensuremath{\left|{#1}\right\rangle}}
\newcommand{\bra}[1]{\ensuremath{\left\langle{#1}\right|}}
\newcommand{\braket}[2]{\ensuremath{\langle{#1}|{#2}\rangle}}
\newcommand{\op}[1]{\ensuremath{\hat{\mathnormal{#1}}}}

\newcommand{\me}[3]{\ensuremath{\langle{#1}|{#2}|{#3}\rangle}}

\DeclareMathOperator{\Tr}{Tr}

\begin{document}

\title{Dynamics of mode entanglement induced by particle-tunneling in the extended Bose-Hubbard dimer model}

\author{Alan J. Barrios}
\affiliation{Instituto de F\'isica, Universidad Nacional Aut\'onoma de M\'exico, Apdo.\ Postal 20-364, 01000, Ciudad de M\'exico, México}

\author{Andrea Vald\'es-Hern\'andez}
\affiliation{Instituto de F\'isica, Universidad Nacional Aut\'onoma de M\'exico, Apdo.\ Postal 20-364, 01000, Ciudad de M\'exico, México}

\author{Francisco J. Sevilla}
\email[]{fjsevilla@fisica.unam.mx}
\thanks{author to whom correspondence should be addressed.}
\affiliation{Instituto de F\'isica, Universidad Nacional Aut\'onoma de M\'exico, Apdo.\ Postal 20-364, 01000, Ciudad de M\'exico, México}


\begin{abstract}
The evolution of mode entanglement is analysed for a system of two indistinguishable bosons with two accessible modes.
Whereas entanglement remains stationary whenever the number of bosons in each mode is left invariant, it exhibits a rich dynamics under the effects of single- and two-particle tunneling. By analysing such effects in paradigmatic families of states, our results provide guidance for the design and control of specific dynamics of mode entanglement, by varying the tunneling transition rates and the preparation of the initial state.
\end{abstract}

\maketitle

\section{Introduction}

The experimental techniques used in ultracold atoms have allowed the study of diverse quantum effects in an extraordinarily controlled manner, as well as the simulation of a variety of theoretical quantum many-body models \cite{BlochNaturePhys2012}. In particular, the dynamics of interacting bosons that hop across two sites (or modes), and models alike, has been relevant to the theoretical \cite{MilburnPRA1997,SmerziPRL1997} and experimental \cite{AndrewsScience1997,SchummNatPhys2005,AnderliniJPhysB2006,GatiJPhysB2007} analysis of Bose-Einstein condensates in a double-well trapping potential. Multiple studies of quantum effects in the double-well model have been advanced related, for example, to: self-trapping \cite{OttavianiPRA2010,CuiPRA2010}, transport \cite{NesterenkoJPB2009}, tunneling manipulation by external fields \cite{ZhangPRA2008}, Josephson tunneling \cite{GatiJPhysB2007,GatiJPhysB2007,MeierPRA2001,FerriniPRA2008}, quantum chaos \cite{KiddPRA2019,ChenPRA2021,CoulletJPhysB2002}, spectral properties \cite{FranzosiPRA2001}, etc.

The further advancement of experimental techniques has allowed the isolation and control of a pair of ultracold atoms in a double well \cite{Sebby-StrabelyPRA2006,Folling2007,TrotzkyScience2008,AnderliniNature2007,MurmannPRL2015}, thus providing suitable systems for implementing different tasks in quantum information science. In this context, mode entanglement acquires particular importance when dealing with identical-particle systems \cite{Dalton2017,Tichy2011,Benatti2020}, and constitutes a useful resource for exploiting quantum correlations, with applications in quantum metrology and teleportation \cite{Benatti2021}. In addition, it has been possible to elucidate the role of mode entanglement in the equilibration process in a system of trapped bosons in optical lattices~\cite{MandelNature2003,DaleyPRL2012,KaufmanScience2016}.
Such systems are characterized by the competition between single-particle tunneling processes of bosons hoping across neighboring lattice sites, and the particles on-site interactions, leading to interesting effects, as the Superfluid-Mott insulator quantum phase transition~\cite{FisherPRB1989,GreinerNature2002}.

Although the atom–atom interactions play an important role in the dynamics of mode entanglement via the effective tunneling 
\cite{NgPRA2003,NgPRA2005,ChizhovPRA2008,DuttaEPB2015,RubeniPRA2017}, the interaction strength between the particles can be turned-off via Feshbach resonances, allowing to focus on the study of the \emph{intrinsic} particle-tunneling dynamics. Tunneling processes have been studied in a system of two interacting identical bosons in a double well potential \cite{ZollnerPRA2008,HunnPRA2013,DobrzynieckiEPJD2016}, in an anharmonic potential  \cite{IshmukhamedovPRA2017}, and in a triple well  \cite{Zhou_NJP2013,WilsmannCommPhys2018}, whereas tunneling dynamics of systems of two fermions are studied in Refs.~\cite{RontaniPRA2013,GharashiPRA2015}. 

Here we focus on a system of two indistinguishable bosons with two accessible modes. We consider the extension of the two-mode Bose-Hubbard Hamiltonian, which takes into consideration the intrinsic effects of single- and two-particle tunneling, to study the dynamics of the mode entanglement of the system. Particular emphasis is placed on those initial states that attain an orthogonal state in a finite amount of time, which sets a temporal scale that characterizes the operation rate of any information processing system \cite{SevillaQRep2021}. Our results contribute to a better understanding of the connection between quantum correlations (entanglement) and the characteristic time scales of evolution from an initial state to a distinguishable one, induced by particle-particle interactions and higher order particle-tunneling processes. Our analysis may also provide guidance for the design and control of specific dynamics of low-dimensional physical systems, with potential applications in quantum information and quantum computation that exploits quantum resources.

The paper is organized as follows: In Sect.~\ref{sectII:System} and Sect.~\ref{sectIII:Entanglement} we introduce, respectively, the system under consideration and the employed quantifier of mode entanglement. In Sect. \ref{subsect:HDiagonal} we
focus on the system's mode entanglement for the class of Hamiltonians that are diagonal in the Fock basis. Although this situation corresponds to the absence of particle tunneling, it allows for a detailed characterization of mode entanglement in the 2-simplex defined by the probability distribution that results when the states are expanded in the Fock basis. In Sect. \ref{sect:Tunneling} we analyze the dynamics of mode entanglement in presence of single- and two-particle tunneling, considering paradigmatic families of initial states, and in Sect. \ref{Tunneling+U} we introduce on-site boson-boson interactions along with  single- or two-particle processes. Finally we present a summary and some concluding remarks in Sect. \ref{sect:Conclusions}.

\section{\label{sectII:System}Evolution under an extended Bose-Hubbard Hamiltonian}
The system under study is described in a three-dimensional Hilbert space $\mathcal{H}$ and is composed of a pair of identical bosons that hop between two sites, $S_{0}$ and $S_{1}$. The system dynamics is dictated by the generic Hamiltonian of the form 
\begin{equation}\label{Generator}
\op{H}=\sum_{\mu,\nu=0,1}\varepsilon_{\mu\nu}^{(1)}\op{a}_{\mu}^{\dagger}\op{a}_{\nu}+\sum_{\lambda,\mu,\nu,\eta=0,1}\varepsilon^{(2)}_{\lambda\mu\nu\eta}
\op { a }^ { \dagger }
_{\lambda}\op{a}^{\dagger}_{\mu}\op{a}_{\nu}\op{a}_{\eta},
\end{equation}
which expresses explicitly one-body and two-body interaction terms, distinguished with superindices (1) and (2), respectively. Here, $\op{a}^{\dagger}_\mu$ and $\op{a}_\mu$ are the standard bosonic creation and annihilation operators, that correspondingly creates or annihilates a boson in site (or mode) $\mu=0,1$. These satisfy the bosonic commutation relations
\begin{equation}\label{relas}
[\op{a}_{\mu},\op{a}^{\dagger}_{\nu}]=\delta_{\mu\nu}, \quad [\op{a}_{\mu},\op{a}_{\nu}]=0,
\end{equation}  
and define the number operator of mode $\mu$ as $\op{n}_{\mu}=\op{a}_{\mu}^{\dagger}\op{a}_{\mu}$. A suitable basis of $\mathcal H$ is thus conformed by the two-mode Fock states
\begin{equation}\label{Fock}
\ket{n}\equiv \ket{2-n,n}=\ket{2-n}_0\otimes \ket{n}_1=\frac{\bigl(\op{a}_{0}^{\dagger}\bigr)^{2-n}}{\sqrt{(2-n)!}}\frac{\bigl(\op{a}_{1}^{\dagger}\bigr)^{n}}{\sqrt{n!}}\ket{\text{vac}},
\end{equation} 
where $\ket{n}_{1}$ denotes the state with $n=0,1,2$ particles in the site $S_{1}$, $\ket{2-n}_{0}$ represents the state with the remaining ($2-n$) particles in the site $S_{0}$, and  $\ket{\text{vac}}$ denotes the two-mode vacuum state.

In writing (\ref{Generator}) we have assumed that only the lowest energy level in each well enters into consideration in the dynamics (two-mode approximation \cite{SmerziPRL1997,MilburnPRA1997}), 
so transitions between different energy levels are excluded from our analysis. The Hamiltonian \eqref{Generator} corresponds to an extension of the standard two-site Bose-Hubbard model, by considering off-site two-body interactions \cite{DobrzynieckiEPJD2016,DobrzynieckiPLA2018,MalJPB2019}. In what follows we will make some symmetry considerations and make clear the physical model that will be studied. 

We start by noticing that the Hermiticity of $\op{H}$ imposes the following conditions on the coefficients $\varepsilon$ in (\ref{Generator}):
\begin{equation}\label{condherm}
\varepsilon^{(1)}_{\mu\nu} = \varepsilon^{(1)}_{\nu\mu},\quad \varepsilon^{(2)}_{\lambda\mu\nu\eta} = \varepsilon^{(2)}_{\eta\nu\mu\lambda}.
\end{equation}
From the first one it follows that
\begin{equation}\label{A2}
\varepsilon^{(1)}_{01} = \varepsilon^{(1)}_{10}\equiv-J^{(1)},
\end{equation}
with $J^{(1)}>0$ the tunneling amplitude between sites. From the second equality in (\ref{condherm}) we have
\begin{subequations}
\begin{align}
 \varepsilon^{(2)}_{0001}+\varepsilon^{(2)}_{0010} =\varepsilon^{(2)}_{1000}+\varepsilon^{(2)}_{0100}\equiv -J^{(2)}_{0},\\
 \varepsilon^{(2)}_{1110}+\varepsilon^{(2)}_{1101} =\varepsilon^{(2)}_{0111}+\varepsilon^{(2)}_{1011}\equiv-J^{(2)}_{1},
\end{align}
\end{subequations}
where $J^{(2)}_{\mu}>0$ denotes the tunneling amplitude of one particle whenever there is one particle at site $\mu$. Further,
\begin{equation}\label{A5}
 \varepsilon^{(2)}_{1100} = \varepsilon^{(2)}_{0011}\equiv-K,
\end{equation}
with $K>0$ representing the two-particle tunneling amplitude between sites. From Eqs. (\ref{A2})-(\ref{A5}) we see that the conditions of Hermiticity (\ref{condherm}) imply that the amplitudes of hopping from mode $0$ to mode $1$ and vice versa are identical, meaning that the tunneling amplitudes are symmetrical.

Finally, we denote with $\varepsilon_{01}$ the interaction energy between sites with one boson each, 
\begin{equation}
\varepsilon_{1001}^{(2)}+\varepsilon_{1010}^{(2)}+\varepsilon_{0101}^{(2)}+\varepsilon_{0110}^{(2)}\equiv\varepsilon_{01},
\end{equation}
and write simply $\varepsilon_{\mu}$ to denote the onsite energy $\varepsilon_{\mu\mu}^{(1)}$, and $U_{\mu}^{(2)}$ to represent the onsite interaction energy between bosons $\varepsilon_{\mu\mu\mu\mu}^{(2)}$. 

With all these considerations, by expanding the sums in Eq. (\ref{Generator}) and rearranging terms, we arrive at
\begin{equation}\label{Generator2b}
\op{H}=\op{H}_{D}+\op{H}_\text{T}+\op{H}_\text{TT},
\end{equation}
with 
\begin{subequations}\label{Hs}
\begin{equation}\label{GeneratorDiagonal}
\op{H}_{D}=\sum_{\mu=0,1}\Bigl[\varepsilon_{\mu}\op{n}_{\mu}+U_{\mu}^{(2)}\op{n}_{\mu}(\op{n}_{\mu}-1)\Bigr]
+\varepsilon_{01}\op {n}_{0}\op{n}_{1},
\end{equation}
\begin{equation}\label{OneParticleTunneling}
\op{H}_\text{T}=-J^{(1)} (\op {a }_ { 1 }^ { \dagger } \op { a }_ { 0 } +\op 
{ a } _ {0 } ^ { \dagger }
\op{a}_{1})-\sum_{\mu=0,1}\Bigl[J^{(2)}_{\mu}\bigl(\op{a}_{1}^{\dagger}\op{n}_{\mu}\op{a}_{0}+\op{a}_{0}^{\dagger}\op{n}_{\mu}\op{a}_{1}\bigr) 
\Bigr],
\end{equation}
and
\begin{equation}
 \op{H}_\text{TT}=-K(\op{a}_{0}^{\dagger}\op{a}_{0}^{\dagger}\op{a}_{1}\op{a}_{1}+\op{a}_{1}^{
\dagger}\op{a} _{1}^{\dagger}\op{a}_{0}\op{a}_{0}).
\end{equation}
\end{subequations}

Now, in the Fock-states basis, \eqref{Generator2b} corresponds to a real symmetric $3\times3$ matrix \begin{equation}
H =
\begin{pmatrix}
H_{00} & H_{01} & H_{02}\\
H_{01} & H_{11} & H_{12}\\
H_{02} & H_{12} & H_{22}
\end{pmatrix},
\label{Matrix1}
\end{equation}
with entries $H_{nm}=\me{n}{\op{H}}{m}$, where  $\ket{n}$ is given in \eqref{Fock}. Explicitly we have
\begin{subequations}
\begin{align}
    H_{00}&=2(\varepsilon_{0}+U_{0}^{(2)}),\\
    H_{01}&=-\sqrt{2}\Bigl(J^{(1)}+J_{0}^{(2)}\Bigr),\\
    H_{02}&=-2K,\\
    H_{11}&=\varepsilon_{0}+\varepsilon_{1}+\varepsilon_{01},\\
    H_{12}&=-\sqrt{2}\Bigl(J^{(1)}+J_{1}^{(2)}\Bigr),\\
    H_{22}&=2(\varepsilon_{1}+U_{1}^{(2)}).
\end{align}
\end{subequations}
From here it follows that the number of parameters in the Hamiltonian can be reduced. It can be further simplified assuming that, except for the energy offset fixed at $\varepsilon_{0}=0$ and $\varepsilon_{1}>0$, the two sites are equivalent and therefore $U\equiv U_{0}^{(2)}=U_{1}^{(2)}$, and $J\equiv J^{(1)}+J_{0}^{(2)}=J^{(1)}+J_{1}^{(2)}$. This reduces the set of parameters to $\varepsilon_{1}$ (the on-site energy in site $S_1$), $\varepsilon_{01}$ (the interaction energy between sites), $U$ (the on-site interaction energy between bosons), and $J$ and $K$ (representing, respectively, the single-particle and the two-particle tunneling amplitudes), and leads finally to the Hamiltonian matrix
\begin{equation}\label{matrixH}
H =
\begin{pmatrix}
2U & -\sqrt{2}J & -2K\\
-\sqrt{2}J & \varepsilon_1+\varepsilon_{01} & -\sqrt{2}J\\
-2K & -\sqrt{2}J & 2(\varepsilon_1+U)
\end{pmatrix}.
\end{equation}


Below we focus our analysis on three cases. The first one, presented in Sect. \ref{subsect:HDiagonal}, prevents the off-diagonal terms in (\ref{matrixH}), so neither single- nor double-particle tunneling are considered. The second case is discussed in Sect. \ref{sect:Tunneling}, in which on-site and off-site particle-particle interaction  energies are turned-off ($U=0$, $\varepsilon_{01}=0$), thus correlated tunneling of the pair of particles results as a consequence of two-body processes rather than of on-site particle-particle interaction, as the process studied in Ref.~\cite{HunnPRA2013}. In Section \ref{Tunneling+U} we analyze as a third case, the effects of the competition between the on-site interaction between particles, and the two different tunneling processes in a separate manner, i.e., we firstly consider $U\neq0$ along with the single-particle tunneling ($J\neq0$) only (corresponding to the standard Bose-Hubbard model), and then we consider $U\neq0$ plus the two-particles tunneling ($K\neq0$).

We consider initial pure states $\ket{\psi(0)}$ in $\mathcal{H}$ and expand them in the basis of eigenvectors $\{\ket{E_i}\}$ of $\hat H$ (which depend implicitly on the Hamiltonian parameters) as \begin{equation}\label{Psi0}
 \ket{\psi(0)}=\sum_{i=1}^{3}\sqrt{r_{i}}e^{\textrm{i}\theta_{i}}\ket{E_{i}},
\end{equation}
where the coefficients $r_i$ are non-negative real numbers satisfying the normalization condition $\sum_{i}r_{i}=1$, and $0\le\theta_{i}\le2\pi$. Each triad $\{r_{i}\}$ thus defines a probability distribution relative to the basis $\{\ket{E_{i}}\}$ and denotes the class of states whose expansion in the energy eigenbasis is given by \eqref{Psi0}. As has been analysed in Ref. \cite{SevillaQRep2021}, the set of distributions $\{r_{i}\}$ is organized into families according to whether or not the class of initial states \eqref{Psi0} associated to $\{r_{i}\}$ reach an orthogonal state in a finite time $\tau$, i.e., the triads $\{r_{i}\}$ are sorted depending on whether or not the state $\ket{\psi(\tau)}=e^{-\textrm{i}\op{H}\tau/\hbar}\ket{\psi(0)}$ is orthogonal to $\ket{\psi(0)}$ for some values of the Hamiltonian's parameters, where the orthogonality condition is prescribed by
\beq\label{orto}
 \braket{\psi(\tau)}{\psi(0)}=\sum_{i=1}^{3}r_{i}e^{\textrm{i}E_{i}\tau/\hbar}=0.
 \eeq

The time evolution of the initial state under the action of $\hat H$ reads \begin{equation}\label{estado}
\ket{\psi(t)}=e^{-\textrm{i}\hat Ht/\hbar}\ket{\psi(0)}=\sum_{i=1}^{3}\sqrt{r_{i}}e^{-\textrm{i}(E_{i}t/\hbar-\theta_{i})}\ket{E_{i}},
\end{equation}
with $E_i$ the eigenvalues of $\op{H}$, i.e., $\hat H\ket{E_i}=E_i \ket{E_i}$. In our subsequent analysis the parameters of (\ref{matrixH}) are chosen so that there are no degeneracies, and therefore the eigenvalues can be ordered according to $E_1<E_2<E_3$. 

\section{\label{sectIII:Entanglement} Mode entanglement}
When considering systems composed of identical particles, the operational notion of entanglement between distinguishable constituents ---based on the possibility of addressing the parties individually---, no longer holds, precisely due to the indistinguishable nature of the components and the concomitant (anti)symmetry of the global state. This has led to different definitions of entanglement in identical-particle composite systems, which allude to the concept of non-local correlations either between particles, or between (addressable) physical
modes (see, e.g., \cite{Benatti2020} and references therein). This latter approach embodies the {\it mode entanglement} as an operationally useful entanglement between {\it distinguishable} occupation states \cite{Dalton2017}. It formally resembles the usual entanglement definition when the state of the composite system is expressed in the Fock representation, and serves as a useful resource in quantum information protocols involving multi-particle systems \cite{Benatti2021}. 

In our two-boson system, we will focus on the dynamics of (localised) mode entanglement; consequently we partitionate the Hilbert space as $\mathcal{H}=\mathcal{H}_{0}\otimes\mathcal{H}_{1}$, with $\mathcal{H}_{\mu}$ the Hilbert space of mode (site) $S_{\mu}$. In the two-mode Fock basis the reduced density matrices $\op{\rho}_{0/1}(t)=\Tr_{1/0}\{\op{\rho}(t)
\}$ (here $\Tr_\mu\{\cdot\}$ denotes the partial trace over states of site $S_{\mu}$) obtained from the density operator $\op{\rho}(t)=\ket{\psi(t)}\bra{\psi(t)}
$ result diagonal and given by
\begin{subequations}\label{rhos}
\begin{eqnarray}
    \op{\rho}_{0}(t)&=&\sum_{n=0}^{2}R_n(t)\ket{2-n}_{0}\bra{2-n}_{0},\\
    \op{\rho}_{1}(t)&=&\sum_{n=0}^{2}R_n(t)\ket{n}_{1}\bra{n}_{1},
\end{eqnarray}
\end{subequations}
with
\begin{align}\label{Concurrence3B}
R_n(t)=&|\braket{\psi(t)}{n}|^2\nonumber\\
=&\sum_{i,j=1}^{3}c_{i,n}c^*_{j,n}\sqrt{r_ir_j}  e^{-\textrm{i}\theta_{ij}} e^{\textrm{i}\omega_{ij}t},
\end{align}
where we have defined
$c_{i,n}=\braket{E_{i}}{n}$, $\theta_{ij}=\theta_{i}-\theta_{j}$, and $\omega_{ij}=(E_{i}-E_{j})/\hbar$. The set $\{R_n(t)\}$ conforms the \emph{time-dependent} probability distribution of the reduced states, satisfying $0\leq R_n(t)\leq 1$, and $\sum_n R_n(t)=1$.  

The operators (\ref{rhos}) represent mixed states in general, and their degree of mixedness allows to quantify the (bipartite) entanglement between modes by means of the \emph{concurrence} \cite{Rungta2001}
\begin{equation}
\label{conc}
C=C_{\mu}(\ket{\psi})=\sqrt{\frac{d}{d-1}\left(1-\Tr\op{\rho}^2_\mu\right)}\leq 1, 
\end{equation}
where $d=\dim\mathcal H_\mu=3$. Since both reduced density matrices \eqref{rhos} have the same eigenvalues, the concurrence is independent of the reduced density matrix chosen. From Eqs. (\ref{rhos}) we get for $C$:
\begin{equation}\label{Concurrence2B}
C(t)=\sqrt{\frac{3}{2}\left(1-\sum_{n=0}^{2}R^2_n(t)\right)},
\end{equation}
which depends implicitly on the set $\{r_{i}\}$ through \eqref{Concurrence3B}, and is a monotonous increasing function of the linear entropy of the probability distribution $\{R_n(t)\}$, $S_{\text{L}}[\{R_n(t)\}]=1-\sum_n R^2_n(t)$. 

As will be clear in the following sections, Hamiltonians \eqref{matrixH} that do not induce particle tunneling between sites, i.e., that are diagonal in the Fock basis, leave the initial concurrence invariant in time. This indicates that diagonal terms alone, which represent interactions between the particles, do not modify the entanglement between the modes. This can be understood by recalling that mode entanglement refers to a correlation between the sites, not between the particles, so interaction terms between the bosons, though may correlate the particles, do not (un)correlate the sites. Accordingly, the tunneling between sites is necessary for the concurrence to exhibit a non-trivial evolution. However, as will be seen in Sect. \ref{sect:Tunneling}, there are non-diagonal Hamiltonians for which a family of initial states exist that in the strong tunneling regime evolve with constant concurrence. Such states (Eqs. (\ref{sslow2}) and (\ref{sslow2b})) have a particular decomposition in terms of the Bell states
\begin{subequations}\label{BellBasis}
\begin{align}
\ket{\Phi^{\pm}}=&\frac{1}{\sqrt{2}}(\ket{0}\pm\ket{2})=\frac{1}{\sqrt{2}}\bigl(\ket{0}_A\otimes\ket{0}_B \pm \ket{1}_A\otimes\ket{1}_B\bigr),\\ \ket{\Psi^{+}}=&\ket{1}=\frac{1}{\sqrt{2}}\bigl(\ket{0}_A\otimes\ket{1}_B + \ket{1}_A\otimes\ket{0}_B\bigr).
\end{align}
\end{subequations}
Here we have resorted to the first quantization formalism and introduced labels $A$ and $B$ to denote one and the other boson, each of which has two accessible states (the modes 0 and 1). Notice that due to the indistinguishable nature of the bosons, the three vectors $\{\ket{\Phi^{\pm}},\ket{\Psi^{+}}\}$ suffice to span the symmetric Hilbert space of the system.

\section{\label{subsect:HDiagonal} Mode entanglement in the absence of particle tunneling}

For the class of Hamiltonians that are diagonal in the Fock basis, and therefore preserve the initial number of particles in each site 
---as is \eqref{matrixH} in the absence of transitions between modes, i.e., $J=K=0$---, we have that $\{\ket{E_i}\}=\{\ket{n}\}$, the initial states are of the form
\begin{equation}\label{Psi0b}
 \ket{\psi(0)}
 =\sum_{n=0}^{2}\sqrt{r_{n}}e^{\textrm{i}\theta_{n}}\ket{n},
\end{equation}
and only one term contributes to the sum in (\ref{Concurrence3B}), so Eq. (\ref{Concurrence2B}) reduces to
\begin{equation}\label{ConcurrenceDiag-1}
C=\sqrt{\frac{3}{2}\left(1-\sum_{i=1}^{3}r_{i}^{2}\right)}.
\end{equation}
The entanglement between modes is thus independent of time and of the initial phases $\theta_{i}$, and depends solely on the energy distribution $\{r_i\}$, reminiscent of \eqref{Concurrence2B}. 

\begin{figure}
    \includegraphics[width=\columnwidth]{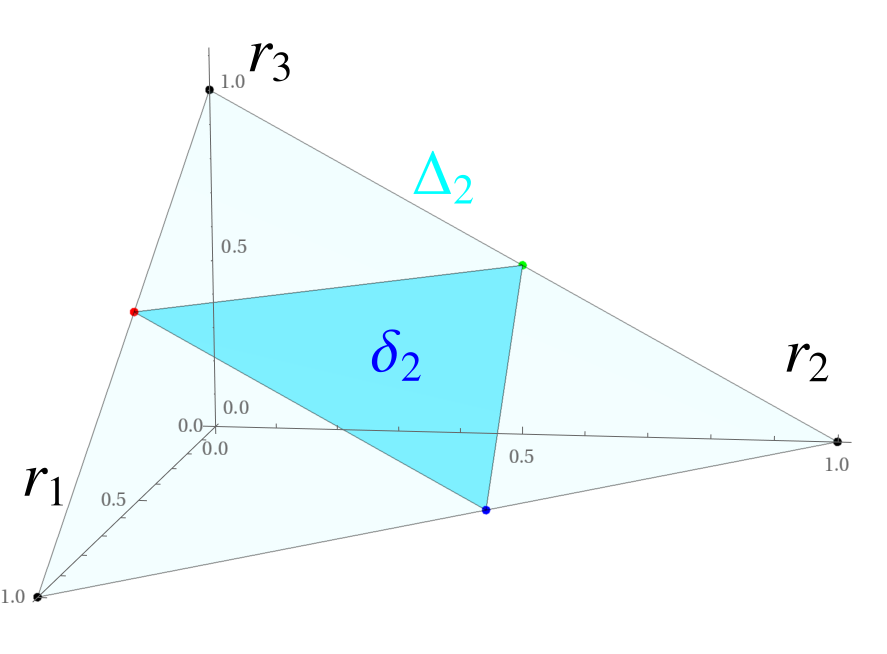}\\
    \includegraphics[width=\columnwidth]{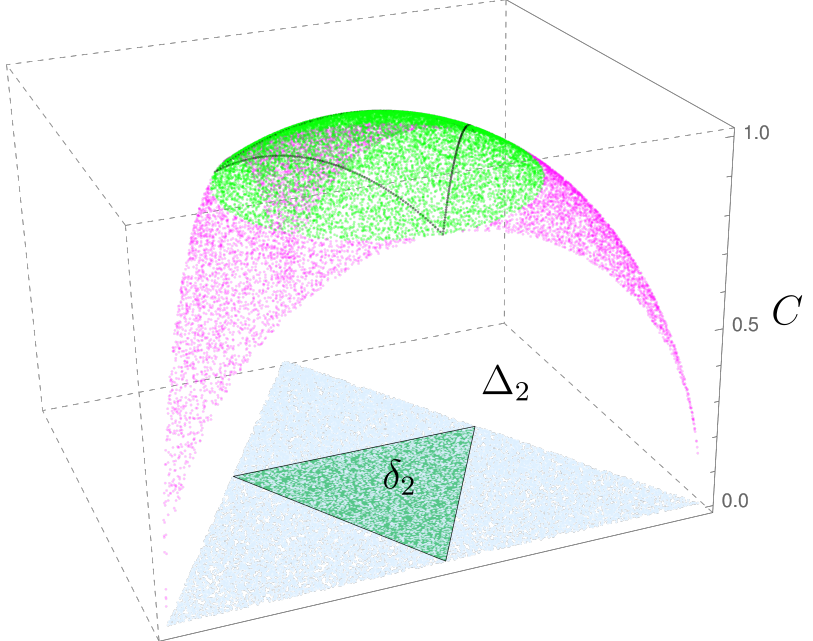}
    \caption{Upper panel shows the 2-simplex $\Delta_{2}$ (big triangle) containing the set of all energy distributions $\{r_i\}$, and the 2-simplex $\delta_2$ (central triangle), containing those triads $\{r_i\}$ for which the corresponding initial state may evolve towards an orthogonal state in a finite time. Bottom panel shows the concurrence $C$ given by Eq. \eqref{ConcurrenceDiag-1} for the families of states determined by the distributions $\{r_{i}\}\in\Delta_{2}$.}
    \label{fig:Simplex}
\end{figure}

The set of all energy distributions $\{r_{i}\}$ is geometrically depicted by the large triangular region shown in the upper panel of Fig. \ref{fig:Simplex}, corresponding to the 2-simplex $\Delta_{2}$. In Ref. \cite{SevillaQRep2021} it was shown that $\Delta_{2}$ splits into
two subsets, 
one corresponding to the central (darker blue) triangle, which is the 2-simplex $\delta_2$, containing all the triads $\{r_i\}$ for which the corresponding initial state \eqref{Psi0} reaches an orthogonal state at a finite time $\tau$ for some value of the Hamiltonian parameters. The other subset is formed from the disconnected (lighter blue) triangles that contain the distributions for which no finite orthogonality time exists, irrespective of the Hamiltonian.

The concurrence corresponding to  distributions in different regions of $\Delta_{2}$ is directly evaluated from 
Eq. \eqref{ConcurrenceDiag-1}, and shown in the bottom panel of Fig. \ref{fig:Simplex}. The vertices of $\Delta_{2}$, identified with the energy distributions $\{1,0,0\}$, $\{0,1,0\}$, and $\{0,0,1\}$, correspond to states \eqref{estado} that are eigenstates of the Hamiltonian (Fock states $\ket{n}$), which are clearly separable in $\mathcal H_0 \otimes \mathcal{H}_1$, and therefore have vanishing entanglement. The distributions represented by the vertices of $\delta_2$, i.e., those corresponding to the triads $\Bigl\{\frac{1}{2},\frac{1}{2},0\Bigr\},\Bigl\{\frac{1}{2},0,\frac{1}{2}\Bigr\},$ and $\Bigl\{0,\frac{1}{2},\frac{1}{2}\Bigr\}$, give rise to qubit (two-level) states whose concurrence settles a threshold for the amount of entanglement of states that evolve towards an orthogonal one in a finite time. Such threshold is given by 
\begin{equation}
C^{\text{q}}=\frac{\sqrt{3}}{2}\approx0.866,
\end{equation}
and delimits the green dome in Fig. \ref{fig:Simplex}.  
Consequently, there are no energy distributions consistent with the orthogonality condition $\langle\psi(\tau)|\psi(0)\rangle=0$ for which $C<C^{\text{q}}$. 

For the distributions lying along the edges of $\delta^2$ ---given by $\Bigl\{\frac{1}{2},r,\frac{1}{2}-r\Bigr\},\Bigl\{r,\frac{1}{2},\frac{1}{2}-r\Bigr\},$ and $
\Bigl\{r,\frac{1}{2}-r,\frac{1}{2}\Bigr\}$ with $0<r<1/2$---, we get 
\begin{equation}
C^{\textrm{edg}}=\sqrt{\frac{3}{4}\Bigl[1+2r(1-2r)\Bigr]},
\end{equation} 
reaching its maximum value $C^{\textrm{edg}}_\text{max}=\sqrt{15}/4$ at $r=1/4$. 
The edges of $\delta_2$ have been projected on the green dome, thus delimiting the values of the concurrence for states characterized by energy distributions inside $\delta_2$. As appreciated in Fig. \ref{fig:Simplex}, the maximum entanglement ($C=1$) is reached at the center of $\delta_2$, corresponding to the equiprobable distribution $r_{1}=r_{2}=r_{3}=1/3$, which is the one that maximizes the linear entropy $S_\text{L}[\{r_i\}]$.

Gathering results we arrive at
\begin{equation}\label{C01fam2}
C^{\textrm{q}}=\frac{\sqrt{3}}{2}\leq C^{\delta_2}\leq 1,
\end{equation}
where $C^{\delta_2}$ represents the concurrence of states whose corresponding $\{r_i\}$ pertain to $\delta_2$, and consequently do reach an orthogonal state in a finite time (for some Hamiltonian). This provides, in particular, a way to determine whether a state of the form (\ref{Psi0}) can evolve, under an appropriate diagonal Hamiltonian, to an orthogonal state by mere inspection of its mode entanglement.

\section{\label{sect:Tunneling}The effects of single- and two-particle tunneling on mode entanglement}

We now proceed to analyze the effects of single- and two particle tunneling in the system mode entanglement. For this, we consider the Hamiltonian 
\begin{equation}\label{matrixHT}
H =
\begin{pmatrix}
0 & -\sqrt{2}J & -2K\\
-\sqrt{2}J & \varepsilon_1& -\sqrt{2}J\\
-2K & -\sqrt{2}J & 2\varepsilon_1
\end{pmatrix},
\end{equation}
which follows from (\ref{matrixH}) after neglecting the terms $\varepsilon_{01}$ and $U$, whence  
is equivalent to
\begin{multline}
\hat{H}=\varepsilon_{1}\hat{n}_{1}
        -J(\hat{a}^{\dagger}_{0}\hat{a}_{1} + \hat{a}^{\dagger}_{1}\hat{a}_{0})\\
        -K\left( \hat{a}^{\dagger}_{0}\hat{a}^{\dagger}_{0}\hat{a}_{1}\hat{a}_{1} + \hat{a}^{\dagger}_{1}\hat{a}^{\dagger}_{1}\hat{a}_{0}\hat{a}_{0}\right).
\label{ND4}
\end{multline}
The Hamiltonian eigenvalues and eigenvectors are computed by exact diagonalization of (\ref{matrixHT}) (see Appendix \ref{sect:SymmetricMatrix}). In particular, the energy spectrum results in 
\begin{subequations}\label{31}
\begin{equation}
E_{k} =\varepsilon_{1}+ \frac{2}{\sqrt{3}}\, \epsilon\cos \left( \frac{\phi+2\pi k}{3} \right),
\label{spectrumg1}
\end{equation}
with $k=1,2,3$,
\begin{equation}\label{spectrumg2}
    \epsilon=\sqrt{\varepsilon^2_1+4(J^2 + K^2)}, 
\end{equation}
and
\begin{equation}
    \phi = \arccos \left(-\frac{12 \sqrt{3} J^2K }{\epsilon^{3}} \right).
\label{spectrumg3}
\end{equation}
\end{subequations}
Notice that whenever either $J$ or $K$ vanish, the eigenvalues yield the equally-spaced spectrum
\begin{equation}\label{spectrumND1}
E_k=\varepsilon_1+(k-2)\epsilon.
\end{equation}

In contrast to the previous case, the tunneling processes induce an intricate time dependence of the mode entanglement, as the phases $\theta_{i}$ of the initial state, the energy spectrum $E_{i}$, and the distribution $\{r_{i}\}$, determine the specific dynamics of $C(t)$. This can be directly evaluated from (\ref{Concurrence2B}), in terms of the dimensionless parameters $J/\varepsilon_{1}$ and $K/\varepsilon_{1}$, after substituting \eqref{spectrumg1} and the corresponding eigenvectors into Eq. \eqref{Concurrence3B}. Since a detailed analysis encompassing a large sample of distributions $\{r_{i}\}$ in $\Delta_{2}$ is impractical, we restrict our analysis to some characteristic  distributions in the simplex $\delta_{2}$ that have been shown to be relevant in the context of quantum speed limit \cite{SevillaQRep2021}. These correspond to the qubit states associated to vertices of $\delta_{2}$, and to qutrit states represented by its center.

\subsection{Two-level states}
We first consider the three vertices of $\delta_{2}$, corresponding to distributions defining families of qubit states. For $\{r_{i}\}=\bigl\{\frac{1}{2},0,\frac{1}{2}\bigr\}$ we focus on the family of initial states 
\begin{subequations}\label{fastslow}
\begin{equation}
\ket{\psi(0)}=\ket{\psi_{\textrm{fast}}}=\frac{1}{\sqrt{2}}(\ket{E_1}+\ket{E_3}),\label{fast}
\end{equation}
whereas for $\{r_{i}\}=\bigl\{\frac{1}{2},\frac{1}{2},0\bigr\}$, $\bigl\{0,\frac{1}{2}, \frac{1}{2}\bigr\}$ we choose, respectively,
\begin{eqnarray}
\ket{\psi(0)}&=&|\psi_{\textrm{slow}}\rangle=\frac{1}{\sqrt{2}}(\ket{E_1}+\ket{E_{2}}),\label{slow}\\
\ket{\psi(0)}&=&|\psi_{\textrm{slow2}}\rangle=\frac{1}{\sqrt{2}}(\ket{E_2}+\ket{E_{3}}).\label{slow2}
\end{eqnarray}
\end{subequations}

Distinctively, states \eqref{fastslow} reach an orthogonal state for the first time at the minimum \emph{orthogonality time} $\tau$, given respectively by 
\begin{subequations}\label{taus}
\begin{eqnarray}
\tau_{\textrm{fast}}&=&\frac{\pi }{\omega_{31}},\\
\tau_{\textrm{slow}}&=&\frac{\pi }{\omega_{21}},\\
\tau_{\textrm{slow2}}&=&\frac{\pi }{\omega_{32}}.
\end{eqnarray}
\end{subequations}
Clearly $\tau_{\textrm{fast}}<\tau_{\textrm{slow}},\tau_{\textrm{slow2}}$, which makes clear the subindices used in states (\ref{fastslow}), stressing that $\ket{\psi_{\textrm{fast}}}$ arrives at an orthogonal state faster than the other two (slower) states. (The ordering relation between $\tau_{\textrm{slow}}$ and $\tau_{\textrm{slow2}}$ varies according to the relative magnitude of the corresponding transition frequencies). The orthogonality time provides a characteristic time scale in the dynamics, which for the states \eqref{fastslow} coincides with the fundamental limit on the rate of quantum dynamics, as has been shown by Levitin and Toffoli \cite{LevitinPRL2009}. The orthogonality time is therefore a useful time scale for analysing the concurrence dynamics in terms of the dimensionless time $t/\tau$. Further, as the evolution of the states \eqref{fastslow} 
is periodic with period $2\tau
$ (see Ref.~\cite{avh2020}), we restrict our analysis to $t/\tau
\in [0,2]$. 

It should be stressed that the states \eqref{fastslow}, as well as the orthogonality times \eqref{taus}, depend on the Hamiltonian's parameters, so the resulting dynamics of the entanglement refers not to a fixed initial state, but rather to a \emph{family} of initial states with a common structure (determined by the corresponding vectors  in (\ref{fastslow})). 

\subsubsection{The family of states $\ket{\psi_{\textrm{fast}}}$}

\begin{figure}[h]     
    \centering
    \subfigure[\label{tinafast1}]
    {
    \includegraphics[width=\columnwidth]{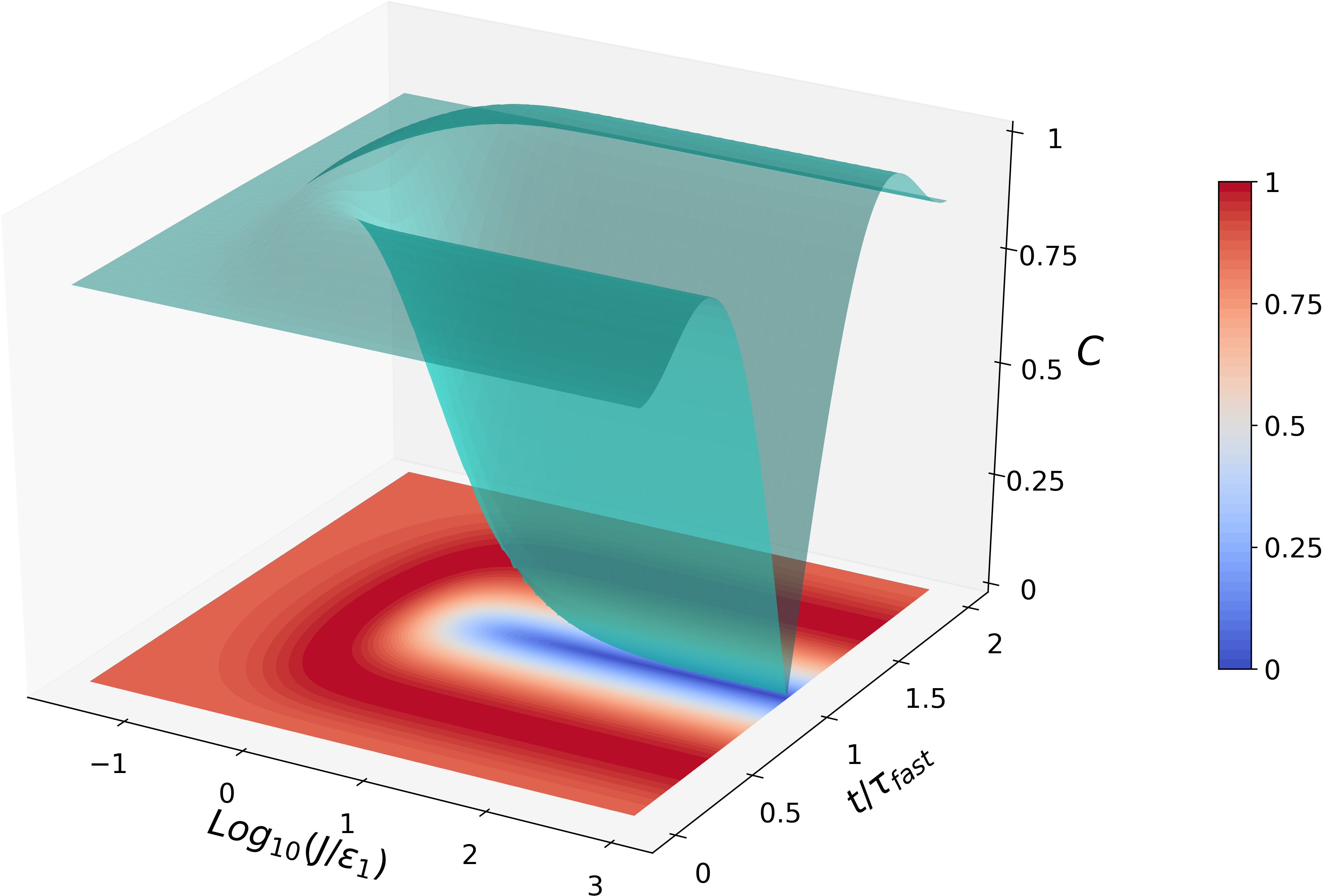}
    }
    \subfigure[\label{tinafast2}]
    {\includegraphics[width=\columnwidth]{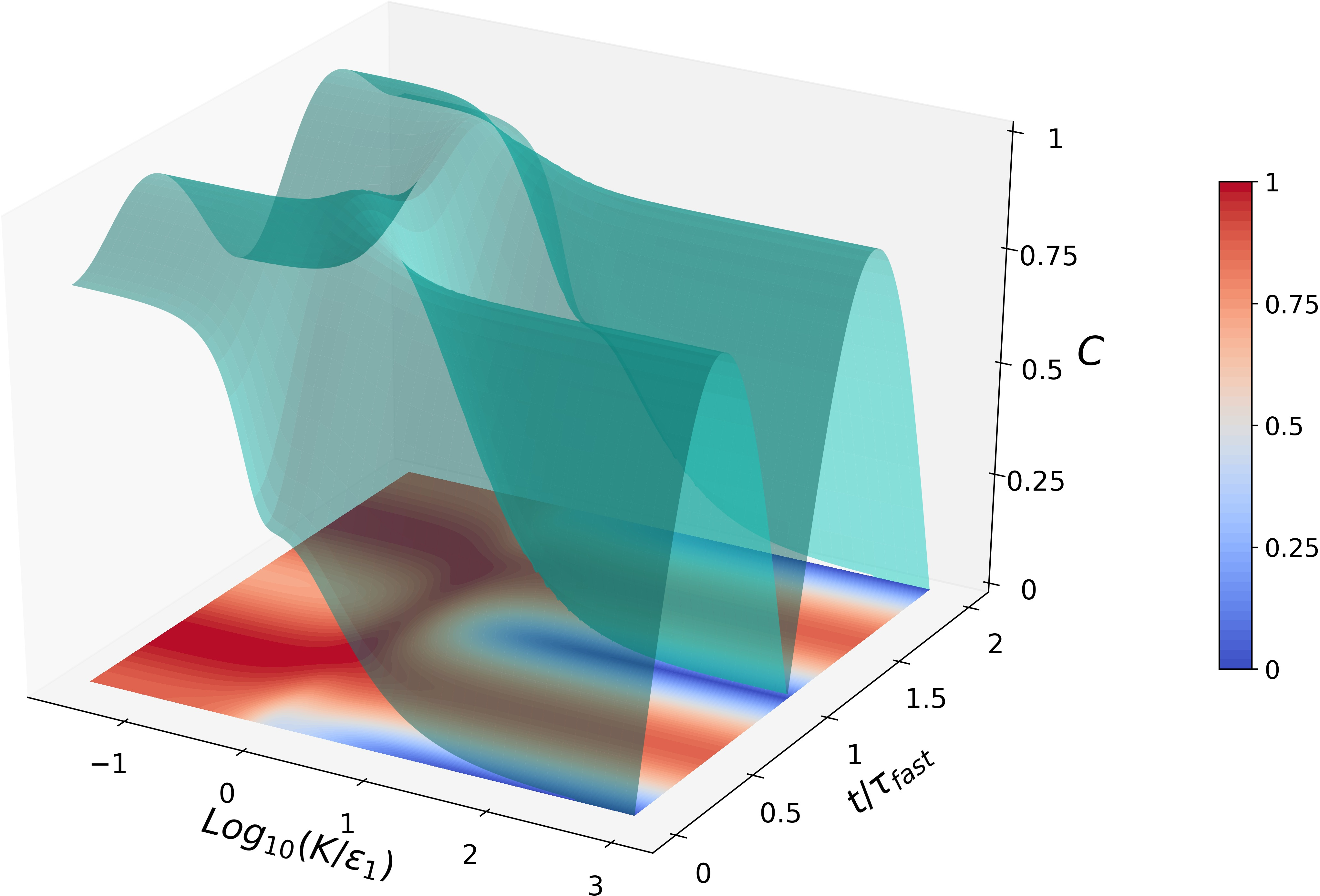}
    }
\caption{Concurrence $C(t)$ for the state $\ket{\psi_\text{fast}(t)}=e^{-\text{i}\op{H}t/\hbar}\ket{\psi_\text{fast}}$ in terms of the dimensionless time $t/\tau_{\textrm{fast}}$ for: a) $K=0$ and varying $J/\varepsilon_1$ (upper panel), and b) $J=\varepsilon_1$ and varying $K/\varepsilon_1$ (bottom panel).}
\label{tinafast}
\end{figure}

The effects of particle tunneling on the concurrence $C(t)$ for the state $\ket{\psi_{\textrm{fast}}(t)}=e^{-\textrm{i}\op{H}t/\hbar}\ket{\psi_\text{fast}}$ are shown in Fig. \ref{tinafast}, as a function of $t/\tau_{\textrm{fast}}$. The case with single-particle tunneling only ($K=0$) is shown in  the upper panel, and the effects of two-particle tunneling, with $J=\varepsilon_{1}$, are shown in the bottom panel. 

\paragraph*{Single-particle tunneling.-} When the system dynamics is driven by single-particle transitions only (Fig. \ref{tinafast1}) the initial states possess all the same entanglement, equal to $C =C^{\textrm{q}}= \sqrt{3}/2\approx 0.86$, independently of the value of $J$. For weak tunneling amplitudes such that $J/\varepsilon_{1}\lesssim 10^{-1}$, the concurrence remains constant throughout the evolution, an expected result since for $K=0$ and $J/\varepsilon_{1} \ll 1$ the Hamiltonian (\ref{matrixHT}) approximates to a diagonal matrix. As $J$ increases and becomes comparable to $\varepsilon_1$, a smooth increase in the correlation between modes takes place
as $t\rightarrow\tau_{\textrm{fast}}$, followed by a decrease in the concurrence as $t>\tau_{\textrm{fast}}$ approaches $2\tau_{\textrm{fast}}$. 

A dramatic change in the evolution of the concurrence occurs in the high-tunneling regime, $J\gg\varepsilon_{1}$. In particular, as $t\rightarrow \tau_{\textrm{fast}}/2$, the entanglement enhances and becomes maximal, 
afterwards decaying abruptly, until at $t=\tau_{\textrm{fast}}$ the system reaches an orthogonal state and the modes disentangle. In this regime the state $\ket{\psi_\text{fast}(t)}$ has, up to a global phase factor, the limiting expression:
\begin{equation}\label{sfast1}
\begin{split}
&\ket{\psi_{\textrm{fast}}(t)}
    \xrightarrow[J\gg\varepsilon_{1}]{}\frac{1}{\sqrt{2}}\bigl(\ket{E_1}+e^{-\text{i}\pi t/\tau_\text{fast}}\ket{E_3}\bigr)\\
    &=
    \cos\Big(\frac{\pi t}{2\tau_{\textrm{fast}}}\Big) \frac{1}{\sqrt 2}\Bigl(  \ket{0}+\ket{2} \Bigr) +  \text{i}\sin\Big(\frac{\pi t}{2\tau_{\textrm{fast}}}\Big)\ket{1}\\
    &=\cos\Big(\frac{\pi t}{2\tau_{\textrm{fast}}}\Big) \ket{\Phi^+} + \text{i}\sin\Big(\frac{\pi t}{2\tau_{\textrm{fast}}}\Big)\ket{\Psi^+},
\end{split}
\end{equation}
while the concurrence reduces to
\begin{equation}
    C(t) \xrightarrow[J\gg\varepsilon_{1}]{} \left\{\frac{3}{2}\left[1- \frac{1}{2} \left( \cos^{4} \frac{\pi t}{2\tau_{\textrm{fast}}}+ 2 \sin^{4}\frac{\pi t}{2\tau_{\textrm{fast}}} \right)\right]\right\}^{1/2}.
\end{equation}
According to Eq. \eqref{sfast1}, the initial state in the strong tunneling regime corresponds to the situation in which both bosons can be found, with equal probability, in either one of the two modes, while at $t=\tau_{\textrm{fast}}$ one of the bosons is found in mode $S_{0}$ whereas the other one is found in mode $S_{1}$, thereby corresponding to the separable Fock state $\text{i}\ket{1}$. 
   
\paragraph*{Two-particle tunneling.-} The effects of two-particle tunneling on the initial states $\ket{\psi_{\textrm{fast}}}$ can be appreciated in Fig. \ref{tinafast2} (for the fixed value $J/\varepsilon_1=1$). In contrast to the previous case, the entanglement of $\ket{\psi_\text{fast}}$ decreases monotonously as 
$K/\varepsilon_{1}$
increases,
vanishing in the limit $K/\varepsilon_{1}\rightarrow\infty$.
In the weak two-particle tunneling regime (highly entangled initial states), we observe that the time dependence of entanglement exhibits small amplitude oscillations, with two maxima at $t\approx \tau_{\textrm{fast}}/2,3\tau_{\textrm{fast}}/2$, and a minimum at $t=\tau_{\textrm{fast}}$. Notice however that even when $K/\varepsilon_{1}\ll J/\varepsilon_{1}=1$, the concurrence dynamics is qualitatively different from that when $K=0$ (compare $C(t)$ at $J/\epsilon_{1}=1$ from Fig. \ref{tinafast1} with $C(t)$ for $K/\varepsilon_{1}\ll1$ of Fig. \ref{tinafast2}), emphasizing that the effects of two-particle tunneling are conspicuous even in the weak regime.
The weak-limit behaviour of $C(t)$ changes for $K/\varepsilon_1\approx 1$, case in which the concurrence does indeed increase as $t$ approximates the orthogonality time. For higher values of $K/\varepsilon_1$, the barely entangled initial states evolve towards highly correlated states, the concurrence reaches a maximum at $t=\tau_{\textrm{fast}}/2$, fall again to its initial value when an orthogonal state is reached, and immediately after, the dynamics is repeated.

For sufficiently large two-particle tunneling amplitudes, such that $K/\varepsilon_{1} \gg J/\varepsilon_1$, we get (up to global phase factors) 
\begin{equation}\label{sfast2}
\begin{split}
&\ket{\psi_{\textrm{fast}}(t)} \xrightarrow[K\gg J,\varepsilon_{1}]{}\frac{1}{\sqrt{2}}\Bigl(\ket{E_1}+e^{-\text{i}\pi t/\tau_\text{fast}}\ket{E_3}\Bigr)
\\
&=\sin\Big(\frac{\pi t}{2\tau_{\textrm{fast}}}\Big) \ket{0} - \text{i} \cos\Big(\frac{\pi t}{2\tau_{\textrm{fast}}}\Big)\ket{2}
\\
&=\sin\Big(\frac{\pi t}{2\tau_{\textrm{fast}}}\Big) \ket{0}_{A}\otimes\ket{0}_{B} - 
\text{i} \cos\Big(\frac{\pi t}{2\tau_{\textrm{fast}}}\Big)\ket{1}_{A}\otimes\ket{1}_{B},
\end{split}
\end{equation}
thus the system periodically oscillates between the two separable Fock states $\ket{0}$ and $\ket{2}$, therefore the bosons tunnel as a pair from site $S_1$ to site $S_0$ and vice versa.
As seen in Fig. \ref{tinafast2}, during the transition from the initial state $\ket{2}$ to the state $\ket{0}$ at $t=\tau_{\textrm{fast}}$, the mode entanglement  increases from $C=0$ up to $C=\sqrt{3}/2$ at $t=\tau_{\textrm{fast}}/2$, and then correlations decrease and vanish when the system reaches a first orthogonal state.
This entanglement-disentanglement process in the strong tunneling regime is characterized by the following structure of the concurrence
\begin{equation}
    C(t) \xrightarrow[K\gg J,\varepsilon_{1}]{} \left\{\frac{3}{8}\left[1-\left(\cos \frac{2 \pi t}{\tau_{\textrm{fast}}}\right)\right]\right\}^{1/2} .
\end{equation}
As in the previous case (upper panel in Fig. \ref{tinafast}), Fig. \ref{tinafast2} shows that the stronger effects on the entanglement occur as the tunneling amplitude increases.

\subsubsection{The family of states $\ket{\psi_\text{slow}}$}

\begin{figure}[h]     
    \centering
    \subfigure[\label{tinaslow1}]
    {
    \includegraphics[width=\columnwidth]{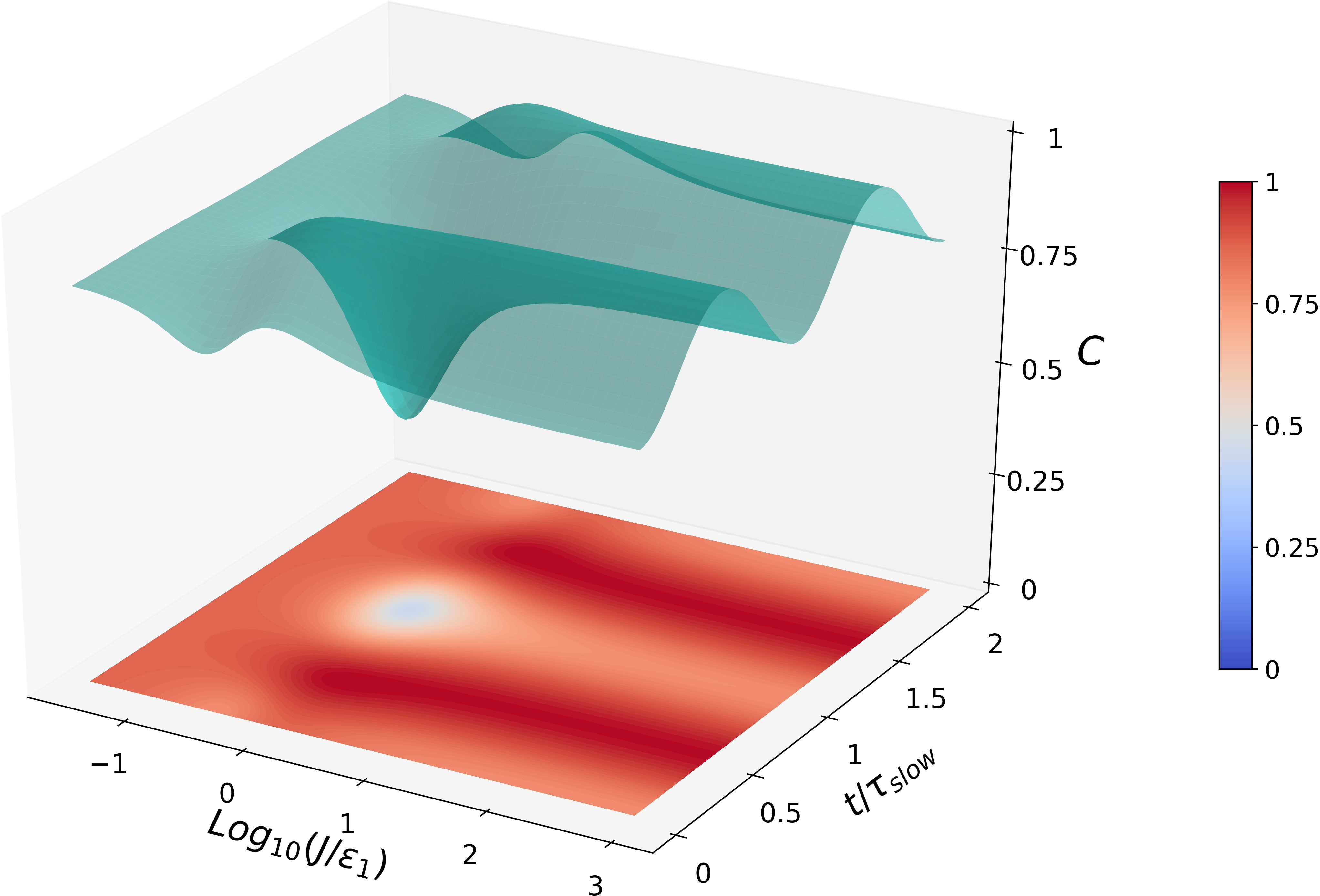}
    }
    \subfigure[\label{tinaslow2}]
    {
    \includegraphics[width=\columnwidth]{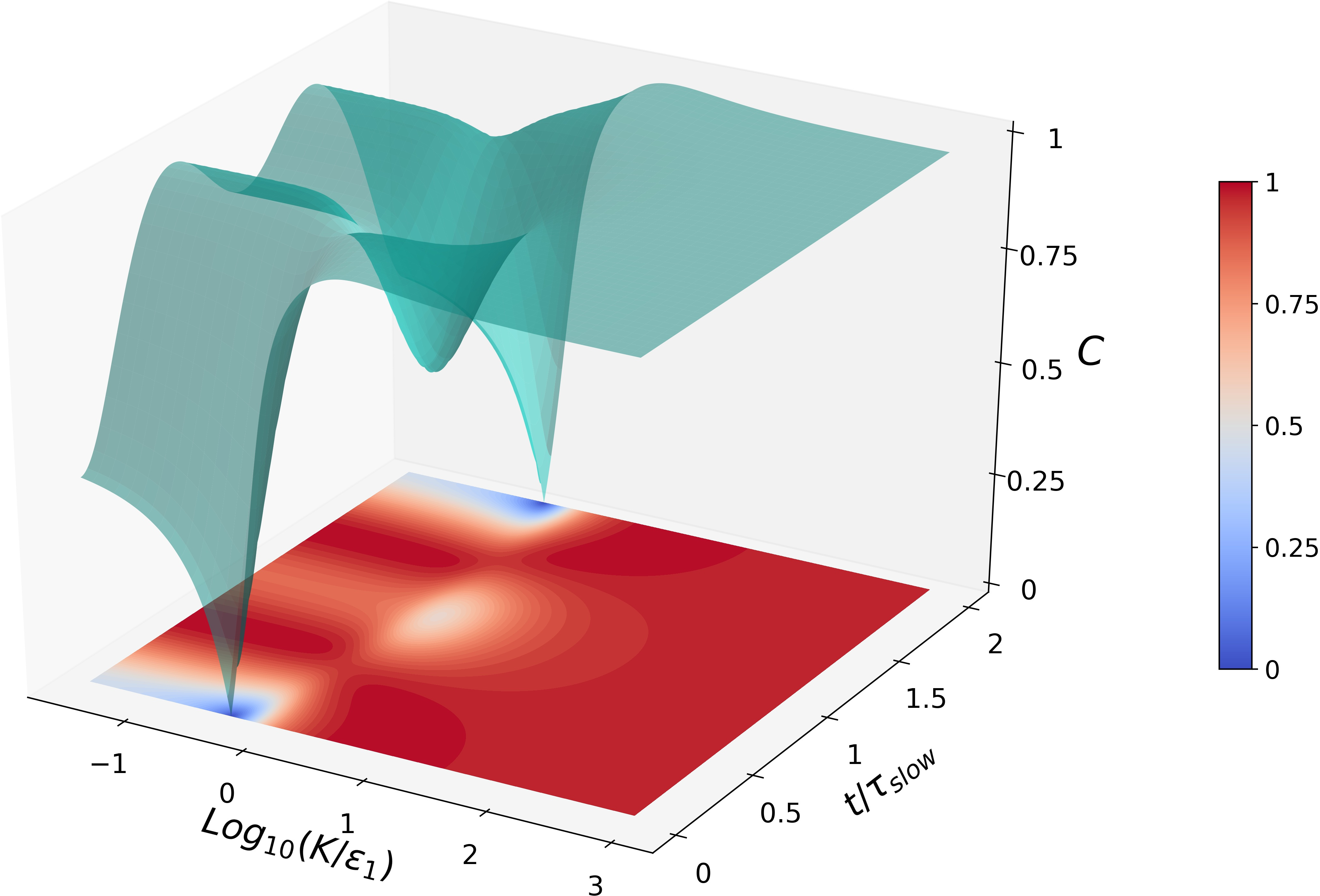}
    }
\caption{Concurrence $C(t)$ for the state $|\psi_{\textrm{slow}}(t)\rangle$ in terms of the dimensionless time $t/\tau_{\textrm{slow}}$ for: a) $K=0$ and varying  $J/\varepsilon_{1}$ (upper panel), and b) $J=\varepsilon_{1}$ and varying  $K/\varepsilon_{1}$ (bottom panel).}
\label{tinaslow}
\end{figure}

When the qubit $\ket{\psi_\text{slow}}$ given in Eq. \eqref{slow} is considered, the time dependence of mode entanglement is shown in Fig. \ref{tinaslow} as a function of the dimensionless time $t/\tau_{\textrm{slow}}$.

\paragraph*{Single-particle tunneling.-} The upper panel of  Fig. \ref{tinaslow} corresponds to the case $K=0$, and exhibits the time dependence of $C(t)$ for different values of $J/\varepsilon_{1}$. 
Notice that the variation of entanglement remains constrained within a relative small range during the evolution for a wide range of values of $J/\varepsilon_1$, except for the global minimum located at $J\approx 0.7 \varepsilon_1$, $t=\tau_\text{slow}$. 
As $J$ is increased above this value, keeping $t=\tau_\text{slow}$, the concurrence grows again saturating to the value $C(\tau_\text{slow})=\sqrt{39}/8\approx0.78$. This means that above this threshold, the entanglement of the state $\ket{\psi_\text{slow}(t)}$ is more robust against the variation of the one-particle tunneling rate. For large values of $J/ \varepsilon_{1}$ the evolved state 
$\ket{\psi_\text{slow}(t)}=e^{-\textrm{i}\hat Ht/\hbar}\ket{\psi_\text{slow}}$ reads (up to a global phase factor), 
\begin{equation}
\begin{split}
    &\ket{\psi_{\textrm{slow}}(t)} \xrightarrow[J\gg \varepsilon_{1}]{}   \frac{1}{\sqrt{2}}\Bigl(\ket{E_1}+e^{-\textrm{i}\pi t/\tau_\text{slow}}\ket{E_2}\Bigr)\\ & = \frac{1}{2} \left[ \left( \frac{1}{\sqrt{2}} - e^{-\textrm{i}\pi t/ \tau_{\textrm{slow}}} \right) \ket{0} \right. +\\  
    &\qquad \qquad \left. \ket{1} + \left( \frac{1}{\sqrt{2}} + e^{-\textrm{i}\pi t/ \tau_{\textrm{slow}}} \right) \ket{2} \right]\\
    &=\frac{1}{2}\left(\ket{\Phi^+}+\ket{\Psi^+}-\sqrt{2}e^{-\textrm{i}\pi t/\tau_\text{slow}}\ket{\Phi^-}\right),
\end{split}
\label{sslow1}
\end{equation}
and the concurrence becomes
\begin{equation}
    C(t) \xrightarrow[J\gg\varepsilon_{1}]{} \left[\frac{3}{16}\left(\frac{17}{4} -\cos{\frac{2\pi t}{\tau_{\textrm{slow}}}} \right)\right]^{1/2} ,
\label{ConcEqSlow}
\end{equation}
whose period is reduced to half the period it had for small values of $J/ \varepsilon_{1}$.

\paragraph*{Two-particle tunneling.-} 
The dynamics of the concurrence when $K/\varepsilon_1$ is varied 
while $J$ is fixed at the value $\varepsilon_1$, is shown at the bottom panel of Fig. \ref{tinaslow}. As in the previous case, the periodicity of the dynamics is twice the orthogonality time. 

When the Hamiltonian includes the two-particle tunneling term so that $K\neq 0$, the entanglement of the initial state $|\psi_{\textrm{slow}}\rangle$ changes abruptly near $K\approx\varepsilon_1$, as can be seen in Fig. \ref{tinaslow2}. Initial states with low entanglement  (corresponding to $K<\varepsilon
_1$) rapidly evolve towards higher entangled states, and maintain a high degree of entanglement during  most part of the period. For initial states which possess a large amount of
entanglement (corresponding to $K>\varepsilon_1$), 
the concurrence does change appreciably only as
$t\rightarrow\tau_{\textrm{slow}}$ for $K/\varepsilon_1\lesssim 10$, whereas in the strong tunneling regime ($K/\varepsilon_1\gg 1$) the concurrence exhibits no evolution at all and remains with the constant value $C=\sqrt{15/16} \approx 0.937$.  This behaviour  can be verified from the structure of the limiting state of $\ket{\psi_\text{slow}(t)}$, i.e.,
\begin{equation}
\begin{split}
 &\ket{\psi_\text{slow}(t)}\xrightarrow[K\gg J]{}\frac{1}{\sqrt{2}}\Bigl(\ket{E_1}+e^{-\text{i}\pi t/\tau_\text{slow}}\ket{E_2}\Bigr) \\
 &=\frac{1}{2} \Bigl(\ket{0}+\ket{2}\Bigr) +   \frac{1}{\sqrt{2}} e^{-\textrm{i}\pi t/ \tau_{\textrm{slow}}} \ket{1}\\
 &=\frac{1}{\sqrt 2}\left(\ket{\Phi^+}+e^{-\text{i}\pi t/\tau_\text{slow}}\ket{\Psi^+}\right)
\end{split}
\label{sslow2}
\end{equation}
for which $R_n$ (see Eq. (\ref{Concurrence3B})) becomes time-independent, resulting in a constant value of the concurrence. 
\subsubsection{The family of states $\ket{\psi_\text{slow2}}$}
\begin{figure}[h]     
    \centering
    \subfigure[\label{tinaslow3}]
    {
    \includegraphics[width=\columnwidth]{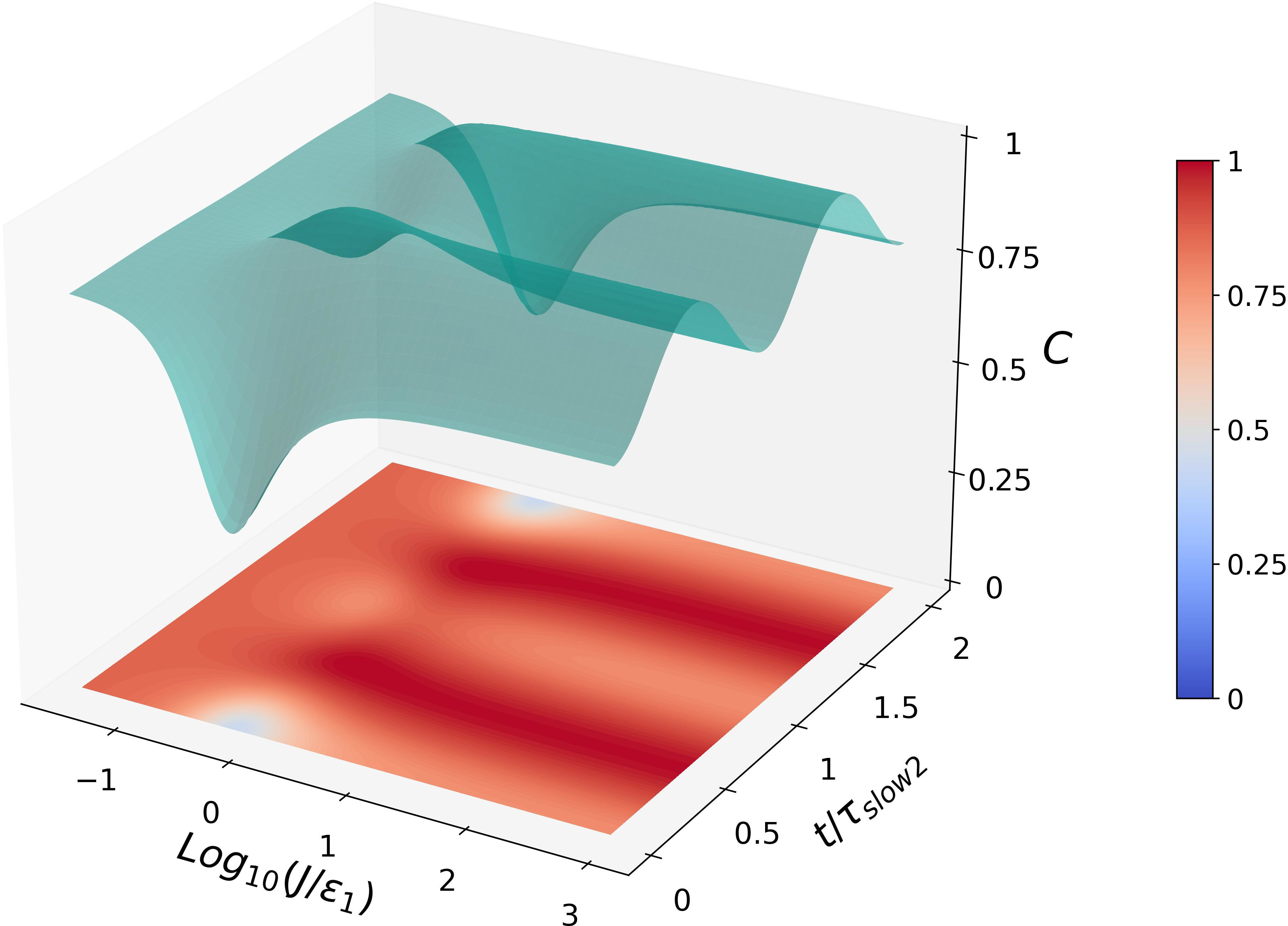}
    }
    \subfigure[\label{tinaslow4}]
    {
    \includegraphics[width=\columnwidth]{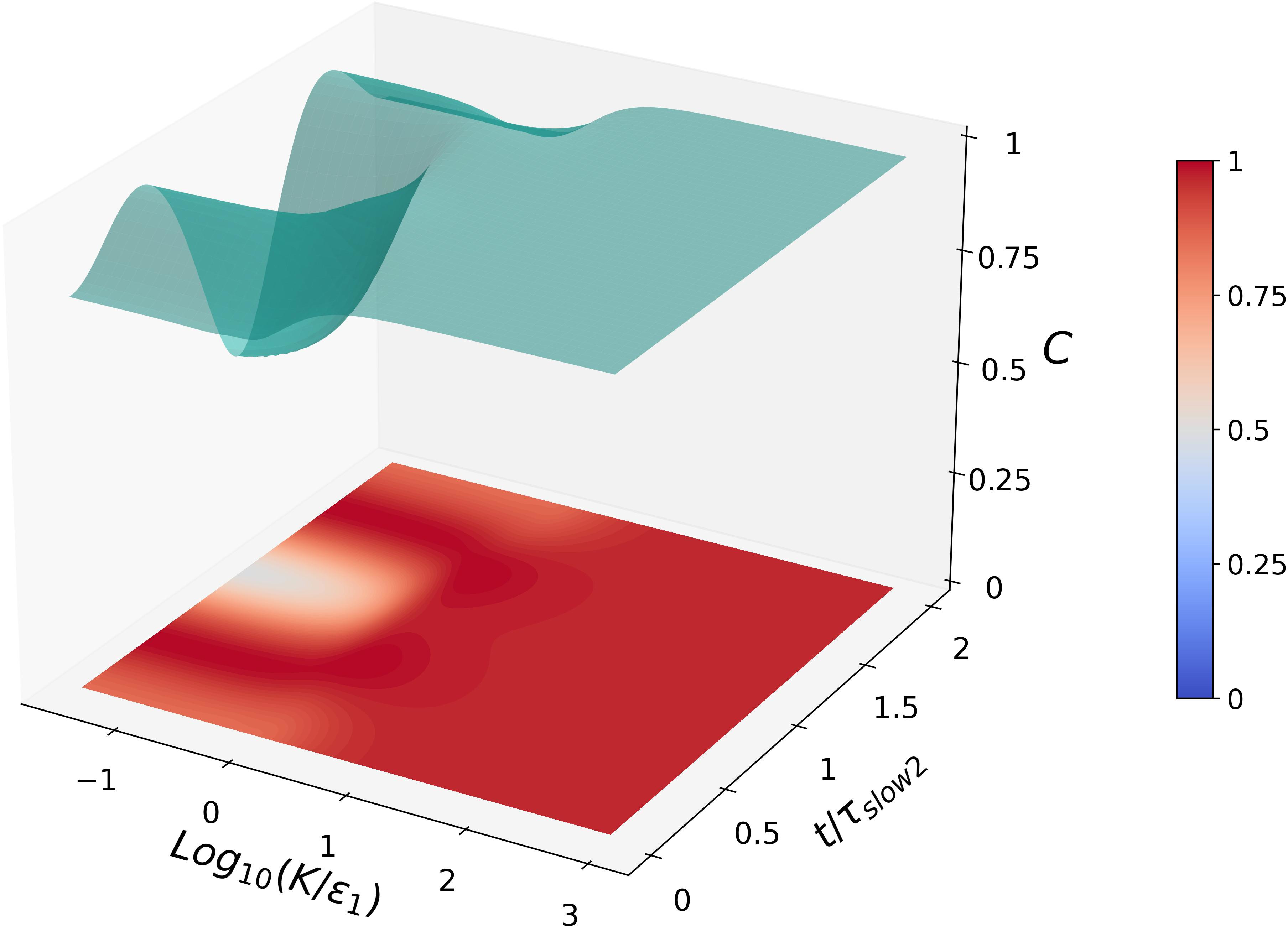}
    }
\caption{Concurrence $C(t)$ for the state $|\psi_{\textrm{slow2}}(t)\rangle$ in terms of the dimensionless time $t/\tau_{\textrm{slow2}}$ for: a) $K=0$ and varying  $J/\varepsilon_{1}$ (upper panel), and b) $J=\varepsilon_{1}$ and varying  $K/\varepsilon_{1}$ (bottom panel).}
\label{tina3qubit}
\end{figure}

The behavior of the concurrence during the evolution of the state $\ket{\psi_{\textrm{slow2}}}$ (Eq.  \eqref{slow2}), is displayed in Fig. \ref{tina3qubit} in terms of the dimensionless parameter  $t/\tau_{\textrm{slow2}}$. As before, the upper panel shows the dynamics of $C$ with $K=0$ as a function of $J/\varepsilon_1$, whereas the bottom panel shows $C(t)$ for fixed $J=\varepsilon_1$ and varying $K$.

\paragraph*{Single-particle tunneling.-} The entanglement between modes in the absence of two-particle tunneling for the state $\ket{\psi_{\textrm{slow2}}}$ exhibits a qualitatively similar behaviour to that corresponding to the state $\ket{\psi_{\textrm{slow}}}$. Indeed, comparison of Figs. \ref{tinaslow1} and \ref{tinaslow3} indicates that appreciable differences between both cases arise only in the regime $J \approx  \varepsilon_{1}$, and manifest in the position and magnitude of the global minima found at $J \approx  0.7\,\varepsilon_{1}$. 
Far from this value of $J$, in the strong one-particle tunneling regime, the concurrence of the state $\ket{\psi_{\textrm{slow2}}(t)}$ replicates the behaviour of the concurrence of the state $\ket{\psi_{\textrm{slow}}(t)}$. This can be verified by noticing that in the limit $J\gg \varepsilon_{1}$ the evolved state 
$\ket{\psi_\text{slow2}(t)}=
e^{-\textrm{i}\hat Ht/\hbar}\ket{\psi_\text{slow2}}$ is, up to a global phase factor,
\begin{equation}
\begin{split}
    &\ket{\psi_{\textrm{slow2}}(t)} \xrightarrow[J\gg \varepsilon_{1}]{}   \frac{1}{\sqrt{2}}\Bigl(\ket{E_2}+e^{-\textrm{i}\pi t/\tau_\text{slow2}}\ket{E_3}\Bigr)\\ & = \frac{1}{2} \left[ \left( \frac{1}{\sqrt{2}}e^{-\textrm{i}\pi t/ \tau_{\textrm{slow2}}} - 1 \right) \ket{0} \right. -\\  
    &\qquad \qquad \left. e^{-\textrm{i}\pi t/ \tau_{\textrm{slow2}}}\ket{1} + \left( \frac{1}{\sqrt{2}}e^{-\textrm{i}\pi t/ \tau_{\textrm{slow2}}} + 1 \right) \ket{2} \right]\\
    &=\frac{1}{2}\Big(e^{-\textrm{i}\pi t/ \tau_{\textrm{slow2}}}\ket{\Phi^+}-\sqrt{2}\ket{\Phi^{-}}-e^{-\textrm{i}\pi t/ \tau_{\textrm{slow2}}}\ket{\Psi^+}\Big),
\end{split}
\label{sslow3}
\end{equation}
which leads to the expression \eqref{ConcEqSlow} for the concurrence as function of time.
\paragraph*{Two-particle tunneling.-} As follows from Fig. \ref{tinaslow4}, by fixing $J=\varepsilon_{1}$ the correlation between modes exhibits two qualitative different dynamics. The first one appears for $K \leq J$ and corresponds to oscillations of  $C(t)$, which attains a minimum value when an orthogonal state is reached. The second type of dynamics emerges for $K > J$, and corresponds in fact to a stationary behaviour of the concurrence (with $C = \sqrt{15/16} \approx 0.937$) as  $K\gg J$, matching what is observed in the case corresponding to the state $\ket{\psi_{\textrm{slow}}}$ in the same limit (cf. Fig. \ref{tinaslow2}). In such strong tunneling regime we get
\begin{equation}
\begin{split}
 &\ket{\psi_\textrm{slow2}(t)}\xrightarrow[K\gg J]{}\frac{1}{\sqrt{2}}\Bigl(\ket{E_2}+e^{-\textrm{i}\pi t/\tau_\textrm{slow2}}\ket{E_3}\Bigr) \\
 &=\frac{1}{2}e^{-\textrm{i}\pi t/\tau_\text{slow2}} \Bigl(\ket{2}-\ket{0}\Bigr) +   \frac{1}{\sqrt{2}} \ket{1}\\
 &=\frac{1}{\sqrt{2}}\Big(
 \ket{\Psi^+}-e^{-\textrm{i}\pi t/\tau_\textrm{slow2}}\ket{\Phi^-}
 \Big).
\end{split}
\label{sslow2b}
\end{equation}
Comparison with Eq. (\ref{sslow2}) shows that the probabilities $R_n$ for both slow states coincide, whence the corresponding mode entanglements coincide in the regime $K\gg J$. The states $\ket{\psi_\textrm{slow}}$ and $\ket{\psi_\textrm{slow2}}$ are examples of a wider class of states that, in the strong tunneling regime, evolve under a non-trivial (non-diagonal) Hamiltonian without changing its mode entanglement, as anticipated at the end of Sect. \ref{sectIII:Entanglement}.

\subsection{The central qutrit of $\delta_2$}\label{casoew}
\begin{figure}[h]     
    \centering
    \subfigure[\label{ew1}]
    {
    \includegraphics[width=\columnwidth]{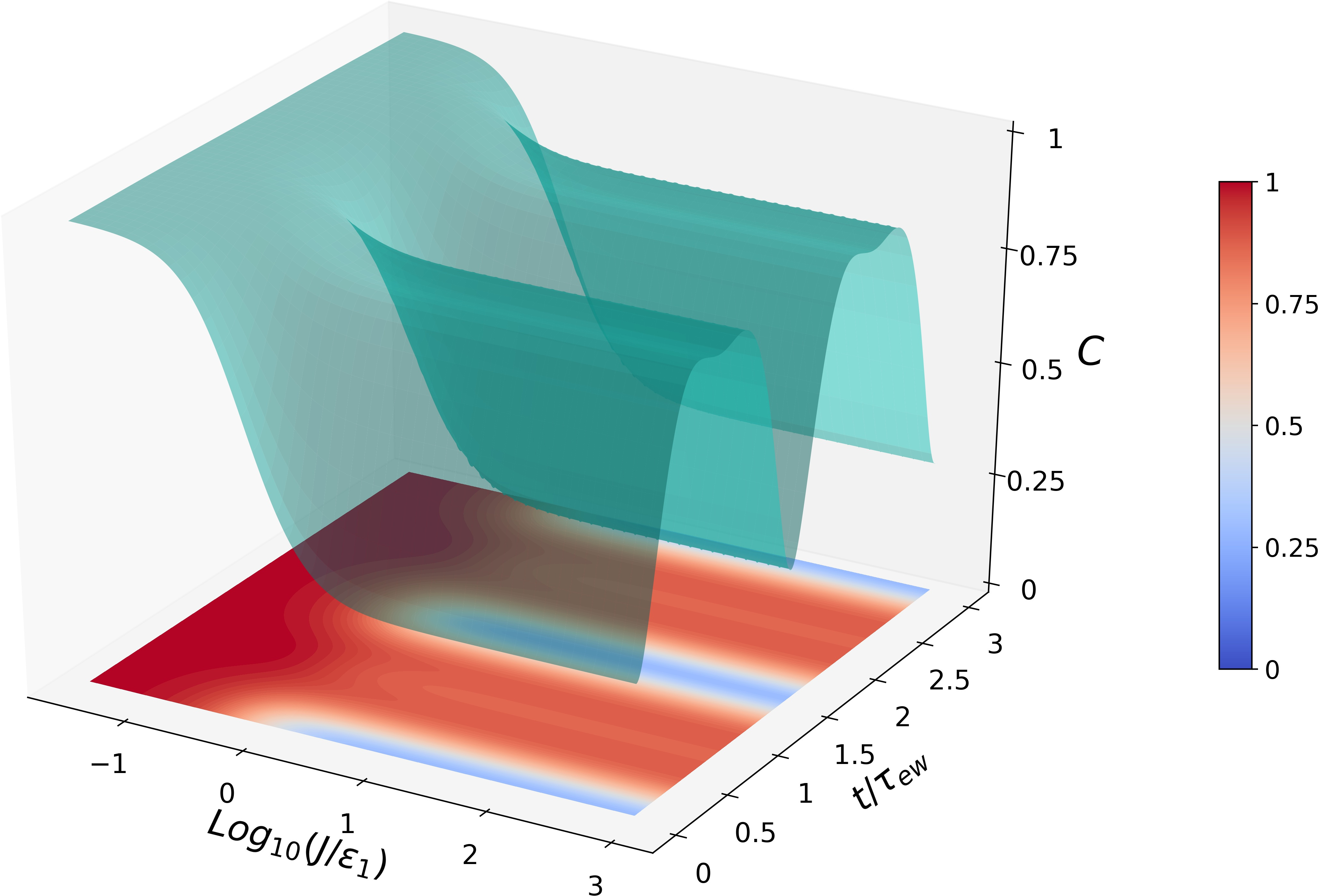}
    }
    \subfigure[\label{ew2}]
    {
    \includegraphics[width=\columnwidth]{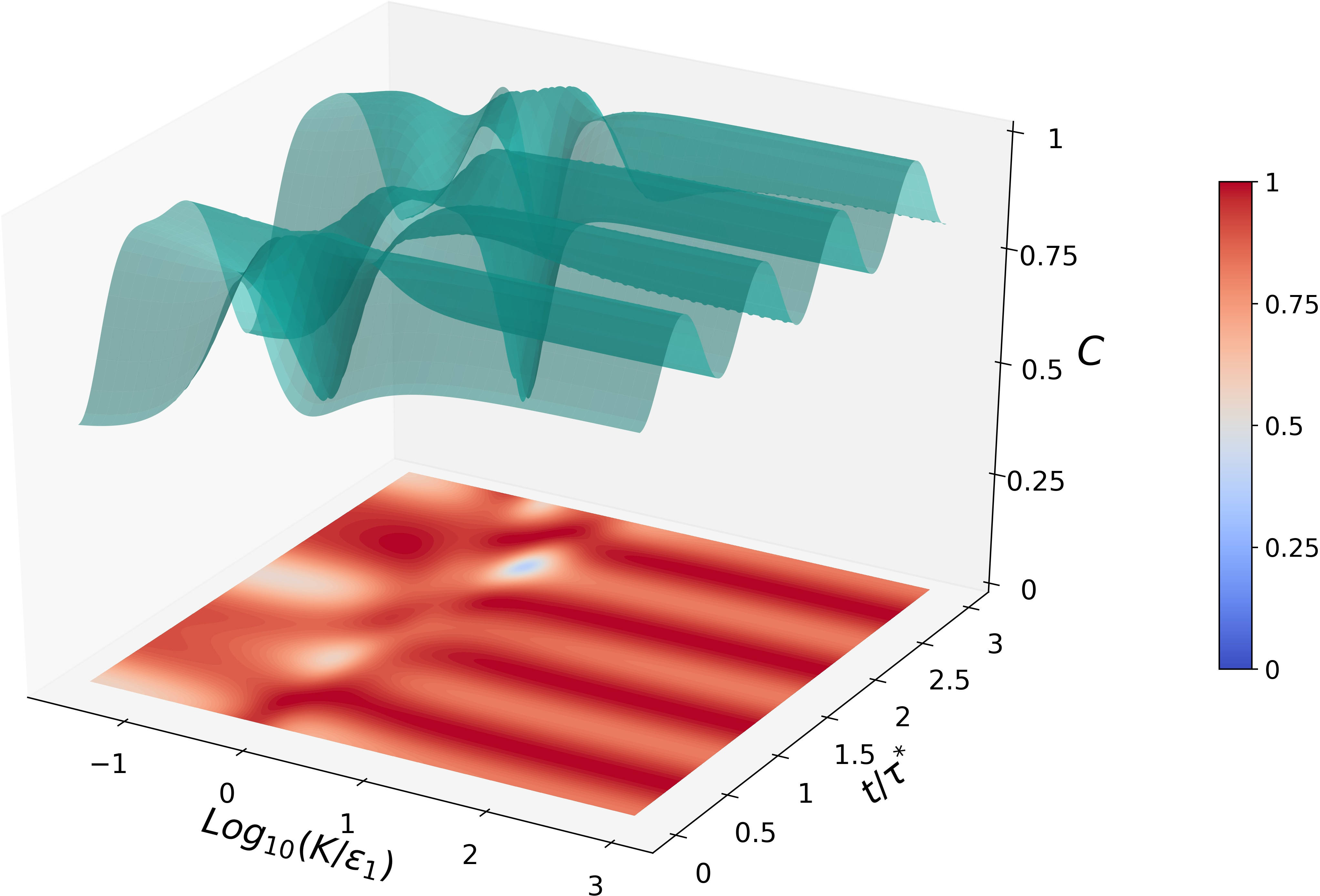}
    }
\caption{Concurrence $C(t)$ for the state $|\psi_{\textrm{ew}}(t)\rangle$ for: a) $K=0$ and varying  $J/\varepsilon_{1}$, in terms of the dimensionless time $t/\tau_{\textrm{ew}}$ (upper panel), and b) $J=\varepsilon_{1}$ and varying  $K/\varepsilon_{1}$, in terms of the dimensionless time $t/\tau^{*}$, with $\tau^{*}= 4\pi/3\omega_{31}$ (bottom panel).}
\label{ew}
\end{figure}

We now consider the center of $\delta_{2}$ corresponding to the distribution $\{r_i\}=\bigl\{\frac{1}{3},\frac{1}{3},\frac{1}{3}\bigr\}$, and focus on the initial state
\begin{equation}
\ket{\psi_{\textrm{ew}}}=\frac{1}{\sqrt{3}}\bigl(\ket{E_1}+\ket{E_2}+\ket{E_3}\bigr).
\end{equation}
The evolution of the concurrence in this case is shown in Fig. \ref{ew} for the two scenarios studied in previous cases, namely $K=0$ (upper panel), and $K\neq0$ with $J=\varepsilon_1$ (bottom panel).

\paragraph*{Single-particle tunneling.-} As stated below Eqs. (\ref{31}), when $K=0$ the energy spectrum is equally-spaced, and given by Eq. (\ref{spectrumND1}). In such case the state $\ket{\psi_\textrm{ew}}$ attains an orthogonal state at  
\beq\label{tauew}
\tau_{\textrm{ew}}=\frac{2\pi\hbar}{3\epsilon}=\frac{2\pi}{3\omega_{21}},\quad \epsilon=\sqrt{\varepsilon^2_1+4J^2},
\eeq
and the period of the dynamics is $3\,\tau_{\textrm{ew}}$ \cite{avh2020}.   
Figure  \ref{ew1} shows the time dependence of the concurrence during this period, in terms of the  dimensionless single-particle tunneling amplitude  $J/\varepsilon_1$. It is observed that the effects of the single-particle tunneling in the state's concurrence are barely appreciable in the weak-tunneling regime and up to $J/\varepsilon_{1} \lesssim 10^{-1}$ ---corresponding to highly-entangled initial states---, and acquire significance for $J/\varepsilon_{1}\gtrsim 1$, corresponding to  low-entangled initial states (with limiting value $C(0) = 1/2\sqrt{3} \approx 0.288$). In this latter case, the mode entanglement rapidly increases and reaches a local maximum (of value $C=\sqrt{37/48} \approx 0.87$) at $t = \tau_{\textrm{ew}}/2$, then slightly reduces and increases again up to the same maximum at $t = \tau_{\textrm{ew}}$, that is, when the first orthogonal state is reached. The concurrence then falls off to its initial value at $t = 3\tau_{\textrm{ew}}/2$, and attains once more the value $C=\sqrt{37/48}$ at $t=2\tau_{\textrm{ew}}$, when the second state that is orthogonal to the initial one, and also to $\ket{\psi_{\textrm{ew}}(t=\tau_{\textrm{ew}})}$, is attained \cite{avh2020}.  

The previously described behaviour extends also to the strong-tunneling regime, in which the evolved state becomes (again, up to a phase factor) 
\begin{equation}
\begin{aligned}
    \ket{\psi_{\textrm{ew}}(t)}& \xrightarrow[J/\varepsilon_1\gg 1]{}   \frac{1}{\sqrt{3}}\Bigl(\ket{E_1}+e^{-2\text{i}\pi t/3\tau_\text{ew}}\ket{E_2}+\\
    &\qquad\qquad\qquad\qquad\qquad\qquad\quad e^{-4\text{i}\pi t/3\tau_\text{ew}}\ket{E_3}\Bigr)\\
    =&\frac{1}{\sqrt{3}} \Big[ \Bigl(-\frac{1}{\sqrt 2}+\cos\frac{2\pi t}{3\tau_{\textrm{ew}}}\Bigr) \ket{0}+   \textrm{i}\sqrt{2} \sin\frac{2\pi t}{3\tau_{\textrm{ew}}}\ket{1}\\
    &\qquad\qquad\qquad\qquad\quad+\Bigl(\frac{1}{\sqrt 2}+\cos\frac{2\pi t}{3\tau_{\textrm{ew}}}\Bigr) \ket{2} \Big]\\
    =&\frac{1}{\sqrt{3}} \Big(\ket{\Phi^-}-  
    \sqrt{2} \cos\frac{2\pi t}{3\tau_{\textrm{ew}}}  \ket{\Phi^+}-\\
    &\qquad\qquad\qquad\qquad\qquad\quad
    \textrm{i}\sqrt{2} \sin\frac{2\pi t}{3\tau_{\textrm{ew}}}\ket{\Psi^+} \Big),
\end{aligned}
\label{sew1}
\end{equation}
with concurrence 
\begin{equation}
    C(t) \xrightarrow[J\gg\varepsilon_{1}]{} \left[\frac{3}{2}- \frac{1}{24} \left( 8\cos \frac{4\pi t}{3\tau_{\textrm{ew}}} + 3\cos \frac{8\pi t}{3\tau_{\textrm{ew}}} + 23 \right)\right]^{1/2},
\end{equation}
whence the period of the state's dynamics is twice the period of the mode-entanglement dynamics.   

\paragraph*{Two-particle tunneling.-} 
For $K \neq 0$ and $J/\varepsilon_{1} = 1$, the parameter $\phi$ in Eq. \eqref{spectrumg3} equals $\pi/2$ only asymptotically for $K\rightarrow \infty$, thus for sufficiently large values of $K$ the energy spectrum results equally-spaced. For such spectrum and for the equally-probable energy distribution, the initial state attains an orthogonal one in a time given by $\tau^{(K\rightarrow \infty)}_{\textrm{ew}}=2\pi/(3\omega_{21})=2\pi/(3\omega_{32})=4\pi/(3\omega_{31})$ \cite{avh2020}. For finite values of $K$, the energy spectrum is no longer equally-spaced, and no orthogonal state is reached for the selected values of the Hamiltonian parameters. In absence of an orthogonality time valid for all $K$, the dynamics of mode entanglement can be analyzed in units of the characteristic time $\tau^{*}=4\pi/3\omega_{31}$, as shown in Fig. \ref{ew2} for fixed $J=\varepsilon_1$. Notice that for sufficiently large $K$ the characteristic time $\tau^*$ coincides with $\tau^{(K\rightarrow \infty)}_{\textrm{ew}}$.

In the regime in which $K\lesssim \varepsilon_1$ the evolution of the concurrence does not exhibit an ordered pattern; it is only when the tunneling amplitude (approximately) exceeds $\varepsilon_1$ that a regular behaviour is observed, that persists in the strong-tunneling regime. It is also for $K> \varepsilon_1$ that the mode entanglement can be enhanced to higher values than those attainable when the parameters are such that $K\lesssim \varepsilon_1$. The observed qualitative change in the dynamics 
points to the fast convergence of the energy spectrum to the equally-spaced one, and the accompanying emergence of a regular evolution between orthogonal states.   
In the regime $K\gg J$, the propagated qutrit evolves (up to a global phase factor) into 
\begin{align}
  \ket{\psi_{\textrm{ew}}(t)}& \xrightarrow[K\gg J]{} \frac{1}{\sqrt{3}}\Bigl(\ket{E_1}+e^{-2\text{i}\pi t/3\tau^*}\ket{E_2}\nonumber\\
    &\qquad\qquad\qquad\qquad\qquad\qquad +e^{-4\text{i}\pi t/3\tau^*}\ket{E_3}\Bigr)\nonumber\\
    =&\frac{1}{\sqrt{3}} \left( \textrm{i}\sqrt{2}\sin\frac{2\pi t}{3\tau^*} \ket{0}  +   
      \ket{1}  +
   \sqrt{2}\cos\frac{2\pi t}{3\tau^*} \ket{2} \right)\nonumber\\
    =&\frac{\sqrt{2}}{\sqrt{3}}\biggl(\textrm{i}\sin\frac{2\pi t}{3\tau^*} \ket{0}_A\ket{0}_B  + 
    \cos\frac{2\pi t}{3\tau^*} \ket{1}_A\ket{1}_B\biggr)\nonumber\\
    &\qquad\qquad\qquad\qquad\qquad\qquad+\frac{1}{\sqrt{3}}\ket{\Psi^+},\label{sew2}
\end{align}
and its concurrence becomes
\begin{equation}
    C(t) \xrightarrow[K \gg J]{} \left(\frac{5}{6}-\frac{1}{6} \cos \frac{8\pi t}{3\tau^*}\right)^{1/2},
\end{equation}
which oscillates between $C=\sqrt{2/3}$ and $C=1$ with period $3\tau^*/4$. The minima of $C$ are located at $t=(3\tau^*/4)n$, whereas its maximum value is attained for $t=(3\tau^*/4)(n+\frac{1}{2})$, with $n=0,1,\dots$. At these latter times the state (\ref{sew2}) reduces to an equally-weighted superposition of Fock states. 
\section{\label{Tunneling+U}Tunneling and particle-particle interactions}

In the previous sections we analyzed the dynamics of mode entanglement in two extreme regimes: Firstly, in the absence of tunneling between sites and $U\neq0$ (Sect. \ref{subsect:HDiagonal}), and later when tunneling processes are present and $U=0$ (Sect. \ref{sect:Tunneling}). In this section we focus on the evolution of mode entanglement considering both, the on-site interaction between particles, and the tunneling of either one or two particles. Specifically, our analysis includes a non-vanishing $U$ and: \emph{a}) the single-particle tunneling process, corresponding to the paradigmatic Bose-Hubbard model, and \emph{b}) two-particle tunneling process. 

\paragraph{The Bose-Hubbard model.-}The Hamiltonian in this case is given by (cf. Eq. (\ref{matrixH}))
\begin{equation}\label{matrixHBH}
H =
\begin{pmatrix}
2U & -\sqrt{2}J & 0\\
-\sqrt{2}J & 0& -\sqrt{2}J\\
0 & -\sqrt{2}J & 2U
\end{pmatrix},
\end{equation}
which takes into account the tunneling of a single particle and the particle-particle on-site interaction. The eigenvectors and eigenvalues of (\ref{matrixHBH}) can be computed from the formulae given in the Appendix \ref{sect:SymmetricMatrix}.

Under the Bose-Hubbard Hamiltonian, the initial states \eqref{Psi0} (with $\theta_{i}=0$ for the sake of simplicity) evolve into 
\begin{multline}\label{psiBH}
\ket{\psi(t)}=\frac{1}{\sqrt{2}}\Biggl(\frac{\sqrt{r_{1}}}{\sqrt{1+\widetilde{U}_{+}^{2}}}-\sqrt{r_{2}}e^{-\textrm{i}\omega_{21}t}+\frac{\sqrt{r_3}e^{-\textrm{i}\omega_{31}t}}{\sqrt{1+\widetilde{U}_{-}^{2}}}\Biggr)\ket{0}\\
+\Biggl(\frac{\sqrt{r_{1}}\widetilde{U}_{+}}{\sqrt{1+\widetilde{U}_{+}^{2}}}+\frac{\sqrt{r_{3}}\widetilde{U}_{-}e^{-\textrm{i}\omega_{31}t}}{\sqrt{1+\widetilde{U}_{-}^{2}}}\Biggr)\ket{1}\\
+\frac{1}{\sqrt{2}}\Biggl(\frac{\sqrt{r_{1}}}{\sqrt{1+\widetilde{U}_{+}^{2}}}+\sqrt{r_{2}}e^{-\textrm{i}\omega_{21}t}+\frac{\sqrt{r_3}e^{-\textrm{i}\omega_{31}t}}{\sqrt{1+\widetilde{U}_{-}^{2}}}\Biggr)\ket{2},
\end{multline}
where we defined
\beq
\widetilde{U}_{\pm}=\frac{U}{2J}\pm\sqrt{1+\biggl(\frac{U}{2J}\biggr)^{2}},
\eeq
and the frequencies that determine the orthogonality times given in \eqref{taus} are given by
\begin{subequations}\label{omegasBH}
\begin{eqnarray}
\omega_{31}&=&\frac{4J}{\hbar}(\widetilde{U}_{+}-\widetilde{U}_{-}),\\
\omega_{21}&=&\frac{2J}{\hbar}\widetilde{U}_{+},\\
\omega_{32}&=&-\frac{2J}{\hbar}\widetilde{U}_{-}.
\end{eqnarray}
\end{subequations}

In the strong boson-boson on-site interaction regime, in which $J/U\ll1$ we find, to first order in $\tilde{J}=J/U$, that the state (\ref{psiBH}) becomes 
\begin{align}\label{psilim}
\ket{\psi(t)}\xrightarrow[\widetilde{J}\ll 1]{}&\frac{1}{\sqrt{2}}\bigl[\sqrt{r_1}\widetilde{J}-e^{-2\textrm{i}t/\tilde{\tau}}(\sqrt{r_2}-\sqrt{r_3})\bigr]\ket{0}\nonumber\\
&\qquad+\bigl(\sqrt{r_1}-\widetilde{J}\sqrt{r_3}e^{-2\textrm{i}t/\tilde{\tau}}\bigr)\ket{1}\nonumber\\
&+\frac{1}{\sqrt{2}}\Bigl[\sqrt{r_1}\widetilde{J}+e^{-2\textrm{i}t/\tilde{\tau}}(\sqrt{r_2}+\sqrt{r_3})\Bigr]\ket{2}\nonumber\\
=&\frac{1}{\sqrt{2}}\bigl(\sqrt{r_1}\widetilde{J}+\sqrt{r_3}e^{-2\textrm{i}t/\tilde{\tau}}\bigr)\ket{\Phi^+}\nonumber\\
&\qquad-\frac{\sqrt{r_2}}{\sqrt{2}}e^{-2\textrm{i}t/\tilde{\tau}}\ket{\Phi^-}\nonumber\\
&+\bigl(\sqrt{r_1}-\widetilde{J}\sqrt{r_3}e^{-2\textrm{i}t/\tilde{\tau}}\bigr)\ket{\Psi^+},
\end{align}
where
\beq
\tilde{\tau}=\hbar/U
\eeq
arises as the natural time scale in this regime. From Eqs. (\ref{taus}) and (\ref{omegasBH}) it is found that in the present approximation, the orthogonality times of the qubit states are related to $\tilde{\tau}$ according to
\begin{subequations}
\begin{align}
\tau_\textrm{fast}\xrightarrow[\widetilde{J}\ll 1]{}&\frac{\pi}{4}\tilde{\tau}\Big(\frac{1}{1+2\tilde{J}^2}\Big)\approx \frac{\pi}{4}\tilde{\tau},\\
\tau_\textrm{slow}\xrightarrow[\widetilde{J}\ll 1]{}&\frac{\pi}{2}\tilde{\tau}\Big(\frac{1}{1+\tilde{J}^2}\Big)\approx \frac{\pi}{2}\tilde{\tau},\\
\tau_\textrm{slow2}\xrightarrow[\widetilde{J}\ll 1]{}&\frac{\pi}{2}\tilde{\tau} \frac{1}{\tilde{J}^2}.\label{slow2lim}
\end{align}
\end{subequations}

\begin{figure}[h]     
    \centering
    \subfigure[\label{BHslow1}]
    {
    \includegraphics[width=\columnwidth]{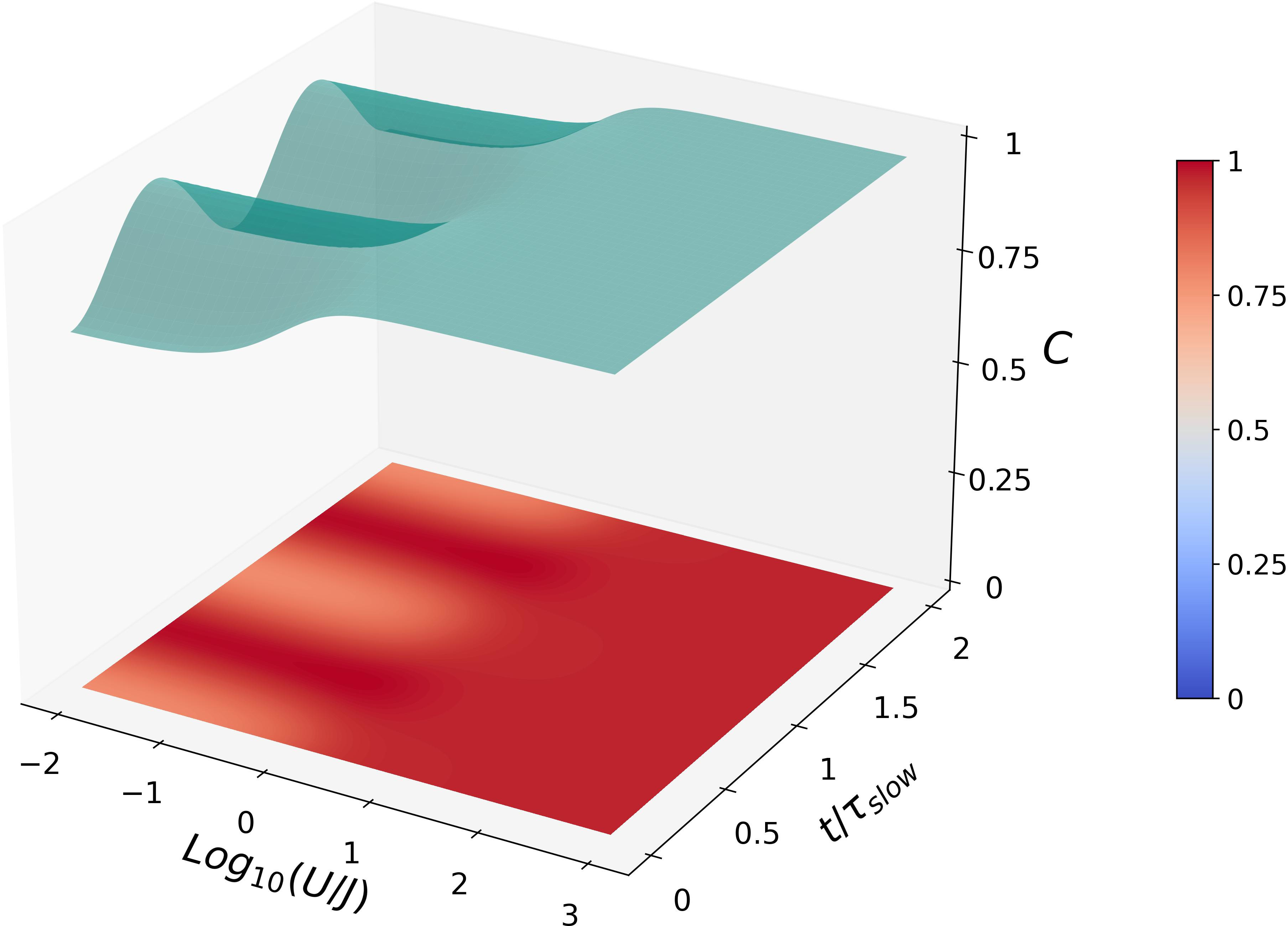}
    }
    \subfigure[\label{BHslow2}]
    {
    \includegraphics[width=\columnwidth]{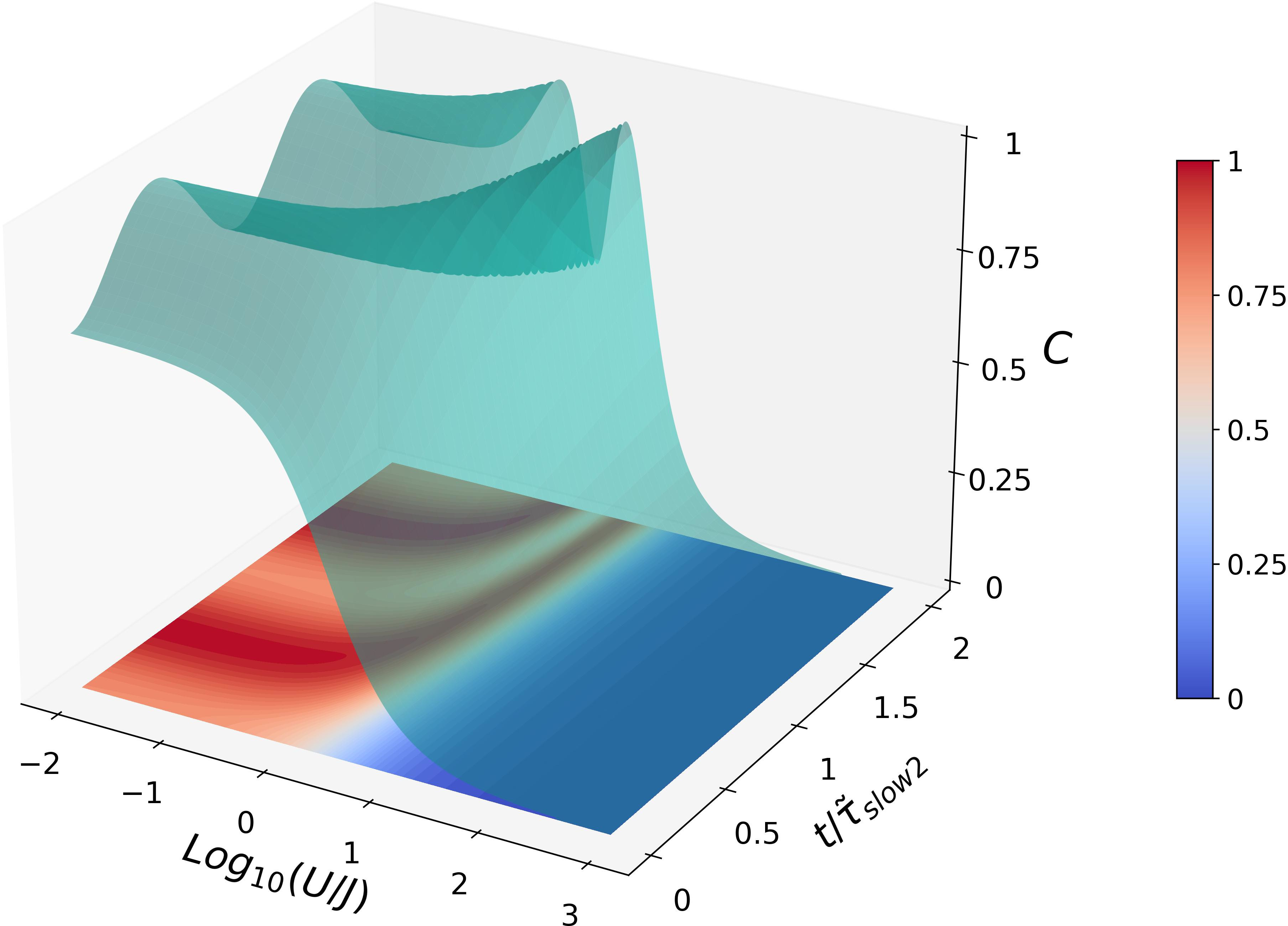}
    }
\caption{Concurrence $C(t)$ as a function of $U/J$ and an appropriate orthogonality time for states: a)$|\psi_{\textrm{slow}}(t)\rangle$ and b) $|\psi_{\textrm{slow2}}(t)\rangle$. In the latter panel $\widetilde{\tau}_\textrm{slow2}= 2\pi/\omega_{31}$.}
\label{BHslow}
\end{figure}

\begin{figure}[h]     
    \centering
    \subfigure[\label{BHfast}]
    {
    \includegraphics[width=\columnwidth]{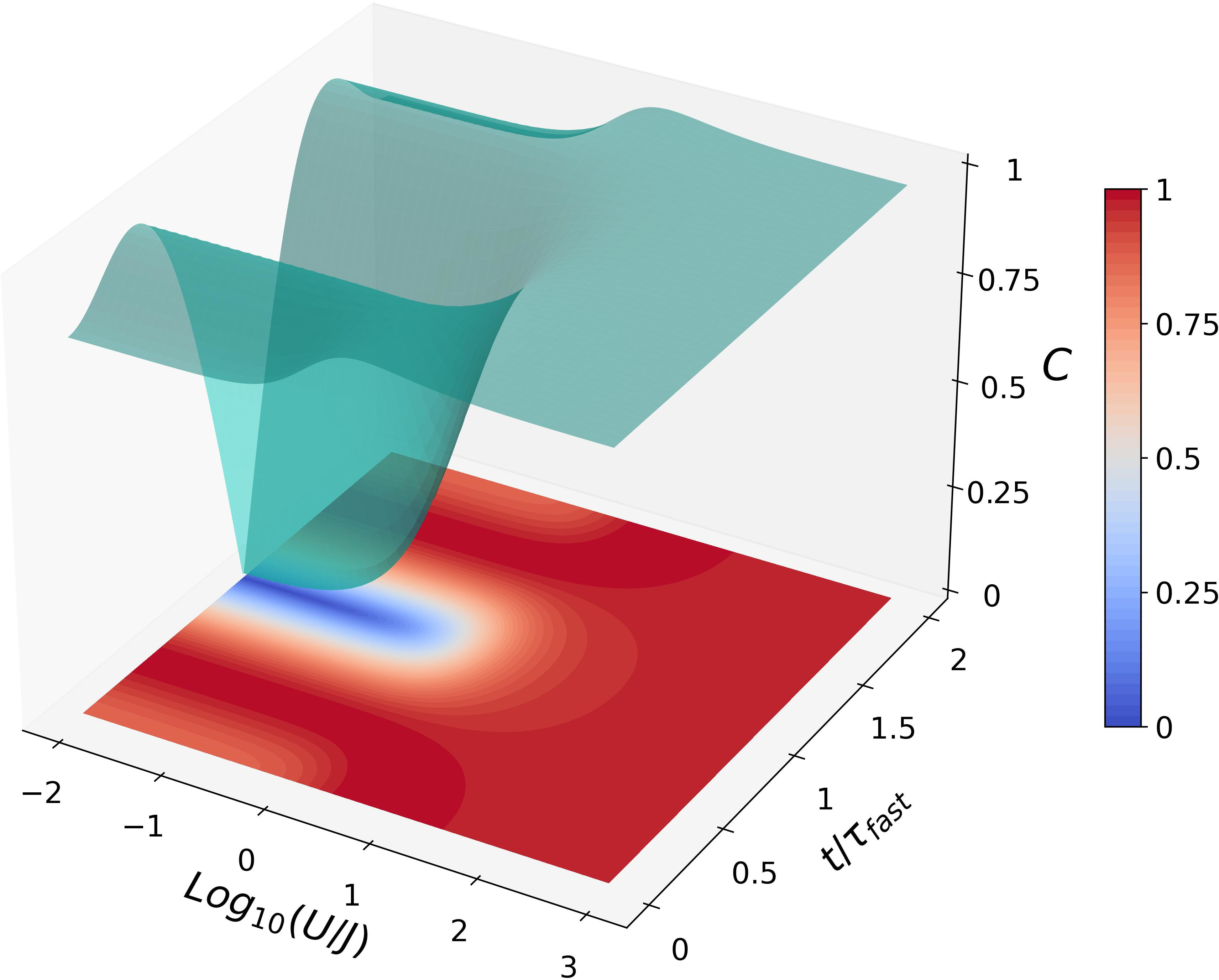}
    }
    \subfigure[\label{BHew}]
    {
    \includegraphics[width=\columnwidth]{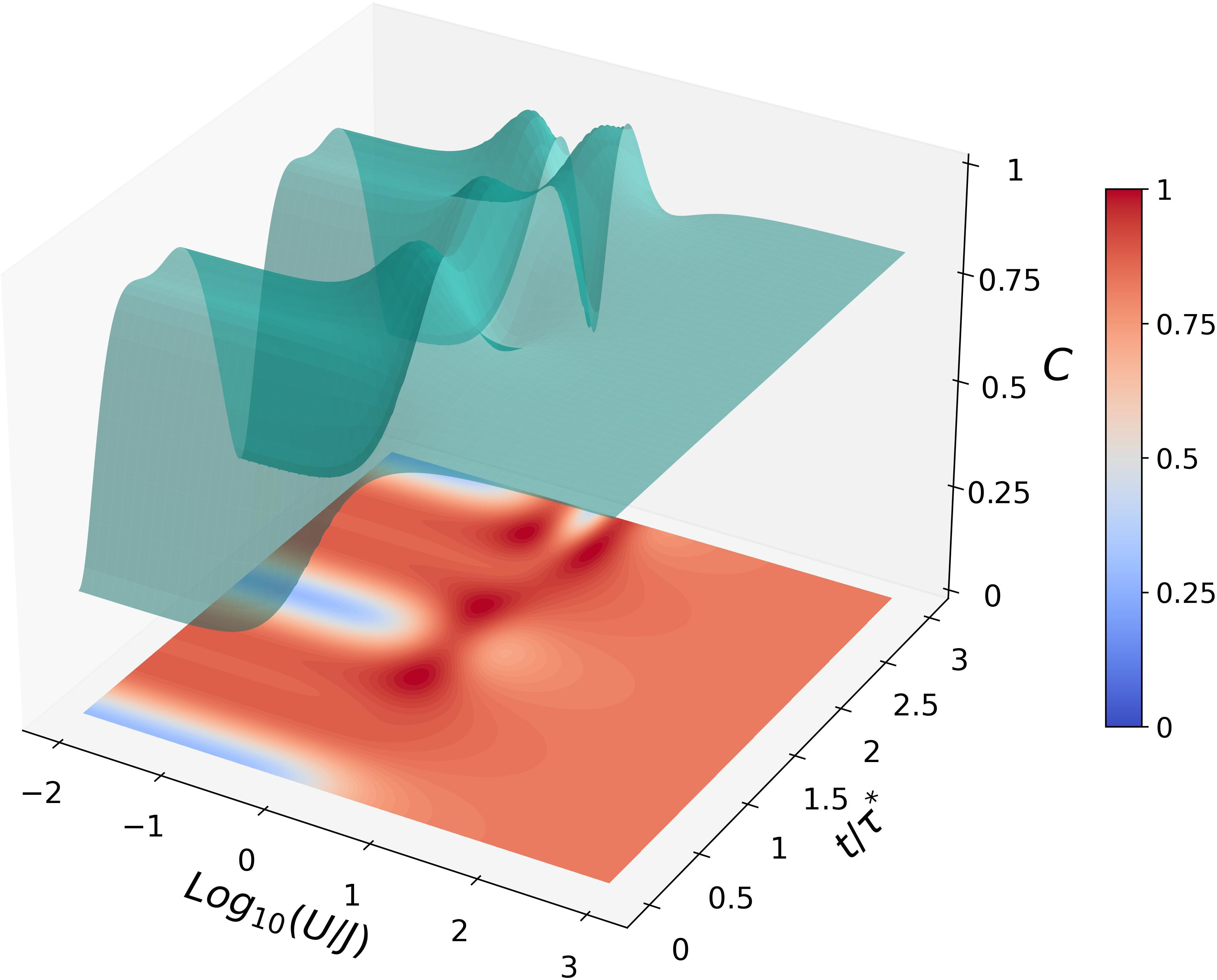}
    }
\caption{Concurrence $C(t)$ as a function of $U/J$ and an appropriate dimensionless time for states: a)$|\psi_{\textrm{fast}}(t)\rangle$ and b) $|\psi_{\textrm{ew}}(t)\rangle$. In the latter panel $\tau^{*}= 4\pi/3\omega_{31}$.}
\label{BHfastew}
\end{figure}

By assigning the appropriate values to the coefficients $\{r_i\}$ in (\ref{psiBH}), the concurrence for the four different states under study can be obtained. Figure \ref{BHslow} shows the dynamics of $C(t)$ for the states $\ket{\psi_\textrm{slow}(t)}$ (upper panel), and $\ket{\psi_\textrm{slow2}(t)}$ (bottom panel), and Figure \ref{BHfastew} exhibits the concurrence for $\ket{\psi_\textrm{fast}(t)}$ (upper panel), and $\ket{\psi_\textrm{ew}(t)}$ (bottom panel). In all cases the concurrence is plotted as a function of $U/J$, and a suitable dimensionless time. For the states
$\ket{\psi_\textrm{slow}(t)}$ and $\ket{\psi_\textrm{fast}(t)}$ such time is determined by the corresponding orthogonality time. For $\ket{\psi_\textrm{slow2}(t)}$, however, choosing $\tau_{\textrm{slow2}}$ introduces singularities in the limit $\widetilde{J}\ll 1$ (see Eq. (\ref{slow2lim})), so in this case the dynamics of $C$ is analysed in terms of the dimensionless time $t/\tilde{\tau}_{\textrm{slow2}}$, with $\tilde{\tau}_{\textrm{slow2}}=2\pi/\omega_{31}$. Finally, for $\tau_{\textrm{ew}}$ we resort to the dimensionless time $t/\tau^*$, with $\tau^*=4\pi/3\omega_{31}$, used in section \ref{casoew}.

As shown in Figs. \ref{BHslow}  and \ref{BHfastew}, the concurrence for all the states crossovers from a noticeable time-varying regime for $U/J\lesssim1$, to a small-varying regime for $U/J\gtrsim1$. From the limit state (\ref{psilim}), we obtain $C(t)$ to first order in $\widetilde{J}$:
\begin{widetext}
\begin{equation}\label{conculim}
C(t)\xrightarrow[\widetilde{J}\ll 1]{}\sqrt{\frac{3}{2}\biggl[1-r_1^2-\frac{1}{2}\bigl(r_2^2+r_3^2\bigr)-3r_2 r_3-\widetilde{J}\sqrt{r_1 r_3}\bigl(5r_2+r_3-4r_1\bigr)\cos(2t/\tilde{\tau})\biggr]}.
\end{equation}
\end{widetext}
For the states $\ket{\psi_\textrm{slow}(t)}$ ($r_1=r_2=\frac{1}{2}, r_3=0$) and $\ket{\psi_\textrm{slow2}(t)}$ ($r_1=0, r_2=r_3=\frac{1}{2}$), the last term in (\ref{conculim}) vanishes, hence the concurrence tends to a constant value at zero order in $\widetilde{J}$. For $\ket{\psi_\textrm{slow}(t)}$ such value is  $\sqrt{\frac{15}{16}}$, whereas for $\ket{\psi_\textrm{slow2}(t)}$ it is zero, as can be appreciated, respectively, in Figs. \ref{BHslow1}, and \ref{BHslow2}. 

For $\ket{\psi_\textrm{fast}(t)}$ ($r_1=r_3=\frac{1}{2}, r_2=0$), Eq. (\ref{conculim}) gives $C(t)\approx\sqrt{\frac{15}{16}}\bigl[1+\frac{3}{5}\widetilde{J}\cos(2t/\tilde{\tau})\bigr]$ and the concurrence reaches the value $\sqrt{\frac{15}{16}}$ as $\widetilde{J}\rightarrow0$. This can be confirmed in Fig. \ref{BHfast} in the regime of high $U/J$.

\paragraph{On-site interaction and two-particles tunneling} We close this section by briefly discussing the case in which the Hamiltonian consists of the on-site interaction and the two-particle tunneling process, so the matrix (\ref{matrixH}) reduces to
\begin{equation}\label{matrixUK}
H =
\begin{pmatrix}
2U & 0 & -2K\\
0 & 0& 0\\
-2K & 0 & 2U
\end{pmatrix}.
\end{equation}

In this case we identify an energy spectrum crossing at $\widetilde{K}\equiv K/U=1$, i.e., while the energy eigenvalues are $0<2(U-K)<2(U+K)$ for $\widetilde{K}<1$, they become $2(U-K)<0<2(U+K)$ for $\widetilde{K}>1$. Since $\tilde{K}=1$ implies a degeneracy in the spectrum, such limit will be excluded from our analysis.     

For $\widetilde{K}<1$ initial states of the form \eqref{Psi0} (again with $\theta_i=0$) evolve, under the Hamiltonian (\ref{matrixUK}), into  \begin{multline}
\ket{\psi(t)}=\frac{1}{\sqrt{2}}e^{-2\textrm{i}t/\tilde{\tau}}\Bigl[\Bigl(\sqrt{r_2}e^{2\textrm{i}\widetilde{K}t/\tilde{\tau}}-\sqrt{r_3}e^{-2\textrm{i}\widetilde{K}t/\tilde{\tau}}\Bigr)\ket{0}+\\
\Bigl(\sqrt{r_2}e^{2\textrm{i}\widetilde{K}t/\tilde{\tau}}+\sqrt{r_3}e^{-2\textrm{i}\widetilde{K}t/\tilde{\tau}}\Bigr)\ket{2}\Bigr]+\sqrt{r_1}\ket{1},
\end{multline}
whereas for $\widetilde{K}>1$ the evolved state reads
\begin{multline}
\ket{\psi(t)}=\frac{1}{\sqrt{2}}e^{-2\textrm{i}t/\tilde{\tau}}\Bigl[\Bigl(\sqrt{r_1}e^{2\textrm{i}\widetilde{K}t/\tilde{\tau}}-\sqrt{r_3}e^{-2\textrm{i}\widetilde{K}t/\tilde{\tau}}\Bigr)\ket{0}+\\
\Bigl(\sqrt{r_1}e^{2\textrm{i}\widetilde{K}t/\tilde{\tau}}+\sqrt{r_3}e^{-2\textrm{i}\widetilde{K}t/\tilde{\tau}}\Bigr)\ket{2}\Bigr]+\sqrt{r_2}\ket{1}.
\end{multline}

With these results the concurrence can be computed in a straightforward manner and is given by
\begin{equation}
C(t)=\begin{cases}
\sqrt{\frac{3}{4}+\frac{3}{2}r_1\bigl(1-\frac{3}{2}r_1\bigr)-3r_2r_3\cos^{2}(4\widetilde{K}t/\tilde{\tau})},\\
\qquad\qquad\qquad\qquad\qquad\qquad\qquad\qquad\textrm{if}\, \widetilde{K}<1, \\
\\
\sqrt{\frac{3}{4}+\frac{3}{2}r_2\bigl(1-\frac{3}{2}r_2\bigr)-3r_1r_3\cos^{2}(4\widetilde{K}t/\tilde{\tau})},\\
\qquad\qquad\qquad\qquad\qquad\qquad\qquad\qquad\textrm{if}\, \widetilde{K}>1.
\end{cases}
\end{equation}
Analytical expressions for the concurrence of the each of the four states considered can thus be obtained. For $\ket{\psi_\textrm{fast}(t)}$:
\begin{equation}\label{UK-Cfast}
C(t)=\begin{cases}
\sqrt{\frac{15}{16}},\quad \textrm{if}\,\, \widetilde{K}<1,\\
\sqrt{\frac{3}{4}\bigl[1-\cos^2( 4\widetilde{K}t/\tilde{\tau})\bigr]}, \quad \textrm{if}\,\, \widetilde{K}>1;
\end{cases}
\end{equation}
for $\ket{\psi_\textrm{slow}(t)}$:
\begin{equation}
C(t)=\sqrt{\frac{15}{16}},\quad \textrm{for}\,\, \widetilde{K}\neq1;
\end{equation}
for $\ket{\psi_\textrm{slow2}(t)}$:
\begin{equation}
C(t)=\begin{cases}
\sqrt{\frac{3}{4}\bigl[1-\cos^2 (4\widetilde{K}t/\tilde{\tau})\bigr]}, \quad \textrm{if}\,\, \widetilde{K}<1,\\
\sqrt{\frac{15}{16}},\quad \textrm{if}\,\, \widetilde{K}>1.
\end{cases}
\end{equation}
At last, for the equally weighted superposition $\ket{\psi_\textrm{ew}(t)}$: 
\begin{equation}
C(t)=\sqrt{1-\frac{1}{3}\cos^2 (4\widetilde{K}t/\tilde{\tau})},\quad \textrm{for}\,\, \widetilde{K}\neq1.
\end{equation}


\section{\label{sect:Conclusions}Conclusions}

We studied the dynamics of mode entanglement in a qutrit system conformed by a pair of two interacting and indistinguishable bosons that can hop between two spatial modes, or sites, paying particular attention to the cases in which: The particles do not tunnel between sites (Sect. \ref{subsect:HDiagonal}), only single- and two-particles tunneling is
allowed (Sect. \ref{sect:Tunneling}), and both, particle interactions and particle tunneling occur (Sect. \ref{Tunneling+U}). In the first case, the concurrence remains constant throughout the evolution and depends only on the (linear entropy of the) energy distribution $\{r_i\}$. This led us to characterize each triad $\{r_i\}$ in the 2-simplex containing all the energy distributions by means of its corresponding concurrence (see Fig. \ref{fig:Simplex}). We thus  established a relation between the amount of entanglement of the initial state and the possibility for it to evolve, in absence of particle tunneling, towards an orthogonal state in a finite time. In particular, we found that a high amount of mode entanglement ($C\geq \sqrt{3}/2$) is necessary in order to reach an orthogonal state, thus providing an entanglement-based criterion to discern whether a given state of the form (\ref{Psi0b}) may evolve into a distinguishable one under an appropriate Hamiltonian. For instance, the initial states $\frac{1}{\sqrt{2}}(\ket{\Phi^{\pm}}+e^{i\theta}\ket{\Psi^{+}})$ have concurrence equal to $\sqrt{15/16}$, thus reach an orthogonal state in a finite time.

When only tunneling amplitudes are present in the Hamiltonian \eqref{matrixH}, the mode entanglement exhibits a rich dynamics, which we analyzed by focusing on four paradigmatic states, three of which ($\ket{\psi_{\textrm{fast}}}$,  $\ket{\psi_{\textrm{slow}}}$ and $\ket{\psi_{\textrm{slow2}}}$) are equally weighted superpositions of two energy eigenstates, and attain an orthogonal state in a finite time. The fourth state ($\ket{\psi_{\textrm{ew}}}$) is an equally weighted superposition of three energy eigenstates, and does not necessarily reach an orthogonal state during the evolution. By varying the Hamiltonian's tunneling parameters $J$ and $K$ these states indeed represent  families of states, whose concurrence time dependence was studied in detail. 

In all cases a qualitative change in the dynamics occurs in passing from the regime in which $J,K<\varepsilon_1$ to that in which $J,K>\varepsilon_1$, and the effects of strong tunneling amplitudes ($K,J\gg\varepsilon_1$) on
$C$ are already present for $K,J\approx 10\varepsilon_1$, so basically increasing the amplitudes beyond this value has no additional effect on the mode entanglement.     

In absence of two-particle tunneling ($K=0$) the concurrence is robust against single-particle hoping in the weak tunneling regime ($J\ll\varepsilon_1$), whereas is more drastically affected in the strong tunneling regime ($J\gg\varepsilon_1$) for $\ket{\psi_{\textrm{ew}}}$ and $\ket{\psi_{\textrm{fast}}}$, and in the intermediate regime ($J\approx\varepsilon_1$) for $\ket{\psi_{\textrm{slow}}}$ and $\ket{\psi_{\textrm{slow2}}}$. Further, comparison of the dynamics of $C(t)$ for the three qubit states
(\ref{fastslow}) suggests that it is the presence of a non-zero component along $\ket{E_2}$ in the initial state what prevents the extreme decay in the concurrence as $t\rightarrow \tau$. In addition, our results show that by an appropriate selection of the initial state, the dynamics induced by the single-particle tunneling can slightly affect the mode entanglement (state $\ket{\psi_{\textrm{slow}}}$), or rather favor an entangling-disentangling dynamics in which highly entangled states evolve into barely (or none) entangled ones and vice versa. (states $\ket{\psi_{\textrm{fast}}}$ and $\ket{\psi_{\textrm{ew}}})$.  As an additional remark, we notice that for the state $\ket{\psi_\text{fast}}$, and  
in the regime with $K=0, J/\varepsilon_1\gg 1$, the mean number of particles in each mode is 1 for all $t$, yet its fluctuations oscillate between 0 (for $t=(2l+1)\tau_\text{fast}$ with $l=0,1,\ldots$) and 1 (for $t=2l\tau_\text{fast}$ with $l=0,1,\ldots$), 
qualitatively following the trend of the concurrence's time dependence. This suggests a connection between the concurrence and the fluctuations of the number of particle in either site, which is left for future analysis. 


When two-particle tunneling processes are considered the dynamics of the mode entanglement is drastically modified, as expected. Interestingly, we observe that an appropriate selection of the value of $K/\varepsilon_1$ may indeed counteract the decrease in the mode entanglement that would occur in the absence of $K$. This occurs in particular for the states $\ket{\psi_{\textrm{ew}}}$, $\ket{\psi_{\textrm{slow}}}$ and $\ket{\psi_{\textrm{slow2}}}$, and can be seen by comparison of Fig. \ref{ew2} in the strong tunneling regime, with the line $J=\varepsilon_1$ ($K=0$) of Fig. \ref{ew1}, and more dramatically by comparison of Fig. \ref{tinaslow2} and Fig. \ref{tinaslow4} in the strong tunneling regime, again with the line $J=\varepsilon_1$ ($K=0$) of Fig. \ref{tinaslow1} and Fig. \ref{tinaslow3}, respectively. Further, from Fig. \ref{tinaslow2} we see that for $K\gg\varepsilon_1$ the mode entanglement saturates its maximum value throughout the evolution, and the state of the pair of bosons decomposes as an equally-weighted superposition of two Bell states (see Eq. (\ref{sslow2})). A similar behaviour is observed when considering the initial state $\ket{\psi_{\textrm{slow2}}}$, as follows from Eq. (\ref{sslow2b}) and can be appreciated in Fig. \ref{tinaslow4}. In contrast, for the state $\ket{\psi_{\textrm{fast}}}$ we observe a different behaviour: for sufficiently high values of $K/\varepsilon_1$ the mode entanglement can indeed be suppressed periodically, as the pair of bosons alternate between the (flipped) states $\ket{00}_{AB}$ and $\ket{11}_{AB}$ (see Eq. (\ref{sfast2})).   

Our findings reveal that states that reach an  orthogonal one evolve into states whose mode entanglement evolves in a regular and periodic manner, in contrast with the disordered dynamics of $C$ for those states that do not comply with the orthogonality condition (\ref{orto}). The relation between the orthogonality time and the characteristic temporal scale of the mode entanglement disclosed here may help to achieve specific amounts of quantum correlations in a controlled way.      

Our analysis concludes addressing the combined effects of particle interactions and particle tunneling on the dynamics of mode entanglement. We focused on the paradigmatic case of the Bose-Hubbard model and a model alike, in which in addition to boson-boson interaction, two-particle tunneling is considered. Both cases allow for a detailed analytical treatment shedding light on the effects of boson-boson interactions as a controlling element of the correlations between sites induced by tunneling.

\acknowledgements
The authors acknowledge support from UNAM-DGAPA PAPIIT-IN110120 (F.J.S. and A.J.B.), and from UNAM-DGAPA PAPIIT-IN113720 (A.V.H.). They also thank the anonymous referees for their valuable criticisms of the manuscript. 

\appendix

\section{\label{sect:SymmetricMatrix}}
Eigenvalues of the Hamiltonian in its generic matrix form \eqref{Matrix1} can be obtained analytically as suggested in \cite{KoppIJMPhy2008}. For this purpose we define 
\begin{subequations}
\begin{align}
    \alpha &= -\Tr H,\\
    \beta  &=\text{det}M_1+\text{det}M_2+\text{det}M_3,\\ 
    \gamma &=-\text{det}H,
\end{align}
\label{MatrixParameter}
\end{subequations}
where $M_1$, $M_2$ and $M_3$ are the three possible $2\times2$ submatrices constructed along the diagonal,
\begin{equation}
M_1 =
\begin{pmatrix}
H_{00} & H_{01} \\
H_{01} & H_{11} 
\end{pmatrix},M_2 =
\begin{pmatrix}
H_{11} & H_{12} \\
H_{12} & H_{22}
\end{pmatrix},M_3 =
\begin{pmatrix}
H_{00} & H_{02} \\
H_{02} & H_{22}
\end{pmatrix}.\nonumber
\end{equation}
In terms of $\alpha$, $\beta$ and $\gamma$, the characteristic polynomial associated to \eqref{Matrix1} is 
\begin{equation}
     E^{3} + \alpha E^{2} + \beta E + \gamma = 0.
     \label{CharacteristicPoly}
\end{equation}
Since we have assumed a non-degenerate spectrum, we look for solutions for which the discriminant $D = p^{3} + q^{2}$ is negative (which implies $p<0$), where $p$ and $q$ are defined according to
\begin{subequations}
\begin{align}
    p &= \dfrac{3\beta - \alpha^{2}}{9},\\
    q &= \dfrac{9\alpha \beta -27\gamma -2\alpha^{3}}{54}.
\end{align}
\end{subequations}
The eigenvalues of $H$ are explicitly given by
\begin{equation}\label{RootH}
    E_{k} = - \frac{\alpha}{3}+2\sqrt{|p|}\cos \left( \frac{\phi+2\pi k}{3} \right),\;k=1,2,3,
\end{equation}
where $\cos\phi = q/|p|^{3/2}$. For $\phi \in [\pi/2, \pi)$ the eigenvalues (\ref{RootH}) are properly ordered according to $E_1<E_2<E_3$, and the transition frequencies read 
\begin{subequations}
\begin{align}
    \omega_{21} &= \frac{2}{\hbar}\sqrt{3|p|}\sin{\frac{\phi}{3}},\\
    \omega_{32} &= \frac{2}{\hbar}\sqrt{3|p|}\cos\left(\frac{\pi + 2\phi}{6}\right),\\
    \omega_{31} &=\omega_{32}+\omega_{21}\\ &=\frac{2}{\hbar}\sqrt{3|p|}\cos\left(\frac{\pi - 2\phi}{6}\right).\nonumber
\end{align}
\label{TransFreq}
\end{subequations}

The eigenvectors of $H$ on the Fock basis are determined by solving the equation $(H - E_{k}\mathbb{I})\Vec{v} = 0$, where $\Vec{v}$ is the vector $(\braket{0}{E_{k}},\braket{1}{E_{k}},\braket{2}{E_{k}})^{\top}$, whose entries $\braket{n}{E_k}$ comply with the normalization condition $\sum_{n} |\braket{E_{k}}{n}|^{2} = 1$. Since these equations have many possible solutions, we consider those for which $H_{01} \neq 0$ and $H_{01} H_{02} - H_{12}(H_{00} - E_{k}) \neq 0$, to get
\begin{subequations}\label{vectors}
\begin{align}
    \braket{E_{k}}{0} = & -\left[ \frac{H_{11} - E_{k}}{H_{01}}A_{k} + \frac{H_{12}}{H_{01}} \right] \braket{E_{k}}{2},\\
    \braket{E_{k}}{1} = & A_{k} \braket{E_{k}}{2},\\
    \braket{E_{k}}{2} = & \left[ \frac{H^{2}_{01} + (H_{11} - E_{k})^{2}}{H^{2}_{01}} A^{2}_{k} + \vphantom{\frac{2 H_{12} (H_{11} - E_{k})}{H^{2}_{01}} A_{k} + \frac{H^{2}_{01} + H^{2}_{12}}{H^{2}_{01}} } \right. \nonumber\\
    & \left. \frac{2 H_{12} (H_{11} - E_{k})}{H^{2}_{01}} A_{k} + \frac{H^{2}_{01} + H^{2}_{12}}{H^{2}_{01}} \right] ^{-1/2}, 
\end{align}
\label{FockProjection}
\end{subequations}
where $A_{k}$ is given by
\begin{equation}
    A_{k} = \frac{(H_{00} - E_{k})(H_{22} - E_{k}) - H^{2}_{02}}{H_{01} H_{02} - H_{12}(H_{00}-E_{k})}.
\end{equation}

In particular, for the Hamiltonian (\ref{matrixHT}) the general expression (\ref{RootH}) becomes equation (\ref{31}), and the restrictions on the solution coefficients (\ref{vectors}) reduce to $J\neq 0$, $E_k\neq 2K$.


\begin{thebibliography}{52}
\expandafter\ifx\csname natexlab\endcsname\relax\def\natexlab#1{#1}\fi
\expandafter\ifx\csname bibnamefont\endcsname\relax
  \def\bibnamefont#1{#1}\fi
\expandafter\ifx\csname bibfnamefont\endcsname\relax
  \def\bibfnamefont#1{#1}\fi
\expandafter\ifx\csname citenamefont\endcsname\relax
  \def\citenamefont#1{#1}\fi
\expandafter\ifx\csname url\endcsname\relax
  \def\url#1{\texttt{#1}}\fi
\expandafter\ifx\csname urlprefix\endcsname\relax\def\urlprefix{URL }\fi
\providecommand{\bibinfo}[2]{#2}
\providecommand{\eprint}[2][]{\url{#2}}

\bibitem[{\citenamefont{Bloch et~al.}(2012)\citenamefont{Bloch, Dalibard, and
  Nascimb{\`e}ne}}]{BlochNaturePhys2012}
\bibinfo{author}{\bibfnamefont{I.}~\bibnamefont{Bloch}},
  \bibinfo{author}{\bibfnamefont{J.}~\bibnamefont{Dalibard}}, \bibnamefont{and}
  \bibinfo{author}{\bibfnamefont{S.}~\bibnamefont{Nascimb{\`e}ne}},
  \bibinfo{journal}{Nature Physics} \textbf{\bibinfo{volume}{8}},
  \bibinfo{pages}{267} (\bibinfo{year}{2012}), ISSN \bibinfo{issn}{1745-2481},
  \urlprefix\url{https://doi.org/10.1038/nphys2259}.

\bibitem[{\citenamefont{Milburn et~al.}(1997)\citenamefont{Milburn, Corney,
  Wright, and Walls}}]{MilburnPRA1997}
\bibinfo{author}{\bibfnamefont{G.~J.} \bibnamefont{Milburn}},
  \bibinfo{author}{\bibfnamefont{J.}~\bibnamefont{Corney}},
  \bibinfo{author}{\bibfnamefont{E.~M.} \bibnamefont{Wright}},
  \bibnamefont{and} \bibinfo{author}{\bibfnamefont{D.~F.} \bibnamefont{Walls}},
  \bibinfo{journal}{Phys. Rev. A} \textbf{\bibinfo{volume}{55}},
  \bibinfo{pages}{4318} (\bibinfo{year}{1997}),
  \urlprefix\url{https://link.aps.org/doi/10.1103/PhysRevA.55.4318}.

\bibitem[{\citenamefont{Smerzi et~al.}(1997)\citenamefont{Smerzi, Fantoni,
  Giovanazzi, and Shenoy}}]{SmerziPRL1997}
\bibinfo{author}{\bibfnamefont{A.}~\bibnamefont{Smerzi}},
  \bibinfo{author}{\bibfnamefont{S.}~\bibnamefont{Fantoni}},
  \bibinfo{author}{\bibfnamefont{S.}~\bibnamefont{Giovanazzi}},
  \bibnamefont{and} \bibinfo{author}{\bibfnamefont{S.~R.}
  \bibnamefont{Shenoy}}, \bibinfo{journal}{Phys. Rev. Lett.}
  \textbf{\bibinfo{volume}{79}}, \bibinfo{pages}{4950} (\bibinfo{year}{1997}),
  \urlprefix\url{https://link.aps.org/doi/10.1103/PhysRevLett.79.4950}.

\bibitem[{\citenamefont{Andrews et~al.}(1997)\citenamefont{Andrews, Townsend,
  Miesner, Durfee, Kurn, and Ketterle}}]{AndrewsScience1997}
\bibinfo{author}{\bibfnamefont{M.~R.} \bibnamefont{Andrews}},
  \bibinfo{author}{\bibfnamefont{C.~G.} \bibnamefont{Townsend}},
  \bibinfo{author}{\bibfnamefont{H.-J.} \bibnamefont{Miesner}},
  \bibinfo{author}{\bibfnamefont{D.~S.} \bibnamefont{Durfee}},
  \bibinfo{author}{\bibfnamefont{D.~M.} \bibnamefont{Kurn}}, \bibnamefont{and}
  \bibinfo{author}{\bibfnamefont{W.}~\bibnamefont{Ketterle}},
  \bibinfo{journal}{Science} \textbf{\bibinfo{volume}{275}},
  \bibinfo{pages}{637} (\bibinfo{year}{1997}),
  \eprint{https://www.science.org/doi/pdf/10.1126/science.275.5300.637},
  \urlprefix\url{https://www.science.org/doi/abs/10.1126/science.275.5300.637}.

\bibitem[{\citenamefont{Schumm et~al.}(2005)\citenamefont{Schumm, Hofferberth,
  Andersson, Wildermuth, Groth, Bar-Joseph, Schmiedmayer, and
  Kr{\"u}ger}}]{SchummNatPhys2005}
\bibinfo{author}{\bibfnamefont{T.}~\bibnamefont{Schumm}},
  \bibinfo{author}{\bibfnamefont{S.}~\bibnamefont{Hofferberth}},
  \bibinfo{author}{\bibfnamefont{L.~M.} \bibnamefont{Andersson}},
  \bibinfo{author}{\bibfnamefont{S.}~\bibnamefont{Wildermuth}},
  \bibinfo{author}{\bibfnamefont{S.}~\bibnamefont{Groth}},
  \bibinfo{author}{\bibfnamefont{I.}~\bibnamefont{Bar-Joseph}},
  \bibinfo{author}{\bibfnamefont{J.}~\bibnamefont{Schmiedmayer}},
  \bibnamefont{and}
  \bibinfo{author}{\bibfnamefont{P.}~\bibnamefont{Kr{\"u}ger}},
  \bibinfo{journal}{Nature Physics} \textbf{\bibinfo{volume}{1}},
  \bibinfo{pages}{57} (\bibinfo{year}{2005}), ISSN \bibinfo{issn}{1745-2481},
  \urlprefix\url{https://doi.org/10.1038/nphys125}.

\bibitem[{\citenamefont{Anderlini et~al.}(2006)\citenamefont{Anderlini,
  Sebby-Strabley, Kruse, Porto, and Phillips}}]{AnderliniJPhysB2006}
\bibinfo{author}{\bibfnamefont{M.}~\bibnamefont{Anderlini}},
  \bibinfo{author}{\bibfnamefont{J.}~\bibnamefont{Sebby-Strabley}},
  \bibinfo{author}{\bibfnamefont{J.}~\bibnamefont{Kruse}},
  \bibinfo{author}{\bibfnamefont{J.~V.} \bibnamefont{Porto}}, \bibnamefont{and}
  \bibinfo{author}{\bibfnamefont{W.~D.} \bibnamefont{Phillips}},
  \bibinfo{journal}{Journal of Physics B: Atomic, Molecular and Optical
  Physics} \textbf{\bibinfo{volume}{39}}, \bibinfo{pages}{S199}
  (\bibinfo{year}{2006}),
  \urlprefix\url{https://doi.org/10.1088/0953-4075/39/10/s19}.

\bibitem[{\citenamefont{Gati and Oberthaler}(2007)}]{GatiJPhysB2007}
\bibinfo{author}{\bibfnamefont{R.}~\bibnamefont{Gati}} \bibnamefont{and}
  \bibinfo{author}{\bibfnamefont{M.~K.} \bibnamefont{Oberthaler}},
  \bibinfo{journal}{Journal of Physics B: Atomic, Molecular and Optical
  Physics} \textbf{\bibinfo{volume}{40}}, \bibinfo{pages}{R61}
  (\bibinfo{year}{2007}),
  \urlprefix\url{https://doi.org/10.1088/0953-4075/40/10/r01}.

\bibitem[{\citenamefont{Ottaviani et~al.}(2010)\citenamefont{Ottaviani,
  Ahufinger, Corbal\'an, and Mompart}}]{OttavianiPRA2010}
\bibinfo{author}{\bibfnamefont{C.}~\bibnamefont{Ottaviani}},
  \bibinfo{author}{\bibfnamefont{V.}~\bibnamefont{Ahufinger}},
  \bibinfo{author}{\bibfnamefont{R.}~\bibnamefont{Corbal\'an}},
  \bibnamefont{and} \bibinfo{author}{\bibfnamefont{J.}~\bibnamefont{Mompart}},
  \bibinfo{journal}{Phys. Rev. A} \textbf{\bibinfo{volume}{81}},
  \bibinfo{pages}{043621} (\bibinfo{year}{2010}),
  \urlprefix\url{https://link.aps.org/doi/10.1103/PhysRevA.81.043621}.

\bibitem[{\citenamefont{Cui et~al.}(2010)\citenamefont{Cui, Wang, and
  Yi}}]{CuiPRA2010}
\bibinfo{author}{\bibfnamefont{B.}~\bibnamefont{Cui}},
  \bibinfo{author}{\bibfnamefont{L.~C.} \bibnamefont{Wang}}, \bibnamefont{and}
  \bibinfo{author}{\bibfnamefont{X.~X.} \bibnamefont{Yi}},
  \bibinfo{journal}{Phys. Rev. A} \textbf{\bibinfo{volume}{82}},
  \bibinfo{pages}{062105} (\bibinfo{year}{2010}),
  \urlprefix\url{https://link.aps.org/doi/10.1103/PhysRevA.82.062105}.

\bibitem[{\citenamefont{Nesterenko et~al.}(2009)\citenamefont{Nesterenko,
  Novikov, Cherny, de~Souza~Cruz, and Suraud}}]{NesterenkoJPB2009}
\bibinfo{author}{\bibfnamefont{V.~O.} \bibnamefont{Nesterenko}},
  \bibinfo{author}{\bibfnamefont{A.~N.} \bibnamefont{Novikov}},
  \bibinfo{author}{\bibfnamefont{A.~Y.} \bibnamefont{Cherny}},
  \bibinfo{author}{\bibfnamefont{F.~F.} \bibnamefont{de~Souza~Cruz}},
  \bibnamefont{and} \bibinfo{author}{\bibfnamefont{E.}~\bibnamefont{Suraud}},
  \bibinfo{journal}{Journal of Physics B: Atomic, Molecular and Optical
  Physics} \textbf{\bibinfo{volume}{42}}, \bibinfo{pages}{235303}
  (\bibinfo{year}{2009}),
  \urlprefix\url{https://doi.org/10.1088/0953-4075/42/23/235303}.

\bibitem[{\citenamefont{Zhang et~al.}(2008)\citenamefont{Zhang, H\"anggi, and
  Gong}}]{ZhangPRA2008}
\bibinfo{author}{\bibfnamefont{Q.}~\bibnamefont{Zhang}},
  \bibinfo{author}{\bibfnamefont{P.}~\bibnamefont{H\"anggi}}, \bibnamefont{and}
  \bibinfo{author}{\bibfnamefont{J.}~\bibnamefont{Gong}},
  \bibinfo{journal}{Phys. Rev. A} \textbf{\bibinfo{volume}{77}},
  \bibinfo{pages}{053607} (\bibinfo{year}{2008}),
  \urlprefix\url{https://link.aps.org/doi/10.1103/PhysRevA.77.053607}.

\bibitem[{\citenamefont{Meier and Zwerger}(2001)}]{MeierPRA2001}
\bibinfo{author}{\bibfnamefont{F.}~\bibnamefont{Meier}} \bibnamefont{and}
  \bibinfo{author}{\bibfnamefont{W.}~\bibnamefont{Zwerger}},
  \bibinfo{journal}{Phys. Rev. A} \textbf{\bibinfo{volume}{64}},
  \bibinfo{pages}{033610} (\bibinfo{year}{2001}),
  \urlprefix\url{https://link.aps.org/doi/10.1103/PhysRevA.64.033610}.

\bibitem[{\citenamefont{Ferrini et~al.}(2008)\citenamefont{Ferrini, Minguzzi,
  and Hekking}}]{FerriniPRA2008}
\bibinfo{author}{\bibfnamefont{G.}~\bibnamefont{Ferrini}},
  \bibinfo{author}{\bibfnamefont{A.}~\bibnamefont{Minguzzi}}, \bibnamefont{and}
  \bibinfo{author}{\bibfnamefont{F.~W.~J.} \bibnamefont{Hekking}},
  \bibinfo{journal}{Phys. Rev. A} \textbf{\bibinfo{volume}{78}},
  \bibinfo{pages}{023606} (\bibinfo{year}{2008}),
  \urlprefix\url{https://link.aps.org/doi/10.1103/PhysRevA.78.023606}.

\bibitem[{\citenamefont{Kidd et~al.}(2019)\citenamefont{Kidd, Olsen, and
  Corney}}]{KiddPRA2019}
\bibinfo{author}{\bibfnamefont{R.~A.} \bibnamefont{Kidd}},
  \bibinfo{author}{\bibfnamefont{M.~K.} \bibnamefont{Olsen}}, \bibnamefont{and}
  \bibinfo{author}{\bibfnamefont{J.~F.} \bibnamefont{Corney}},
  \bibinfo{journal}{Phys. Rev. A} \textbf{\bibinfo{volume}{100}},
  \bibinfo{pages}{013625} (\bibinfo{year}{2019}),
  \urlprefix\url{https://link.aps.org/doi/10.1103/PhysRevA.100.013625}.

\bibitem[{\citenamefont{Chen et~al.}(2021)\citenamefont{Chen, Keiler, Xianlong,
  and Schmelcher}}]{ChenPRA2021}
\bibinfo{author}{\bibfnamefont{J.}~\bibnamefont{Chen}},
  \bibinfo{author}{\bibfnamefont{K.}~\bibnamefont{Keiler}},
  \bibinfo{author}{\bibfnamefont{G.}~\bibnamefont{Xianlong}}, \bibnamefont{and}
  \bibinfo{author}{\bibfnamefont{P.}~\bibnamefont{Schmelcher}},
  \bibinfo{journal}{Phys. Rev. A} \textbf{\bibinfo{volume}{104}},
  \bibinfo{pages}{033315} (\bibinfo{year}{2021}),
  \urlprefix\url{https://link.aps.org/doi/10.1103/PhysRevA.104.033315}.

\bibitem[{\citenamefont{Coullet and Vandenberghe}(2002)}]{CoulletJPhysB2002}
\bibinfo{author}{\bibfnamefont{P.}~\bibnamefont{Coullet}} \bibnamefont{and}
  \bibinfo{author}{\bibfnamefont{N.}~\bibnamefont{Vandenberghe}},
  \bibinfo{journal}{Journal of Physics B: Atomic, Molecular and Optical
  Physics} \textbf{\bibinfo{volume}{35}}, \bibinfo{pages}{1593}
  (\bibinfo{year}{2002}),
  \urlprefix\url{https://doi.org/10.1088/0953-4075/35/6/312}.

\bibitem[{\citenamefont{Franzosi and Penna}(2001)}]{FranzosiPRA2001}
\bibinfo{author}{\bibfnamefont{R.}~\bibnamefont{Franzosi}} \bibnamefont{and}
  \bibinfo{author}{\bibfnamefont{V.}~\bibnamefont{Penna}},
  \bibinfo{journal}{Phys. Rev. A} \textbf{\bibinfo{volume}{63}},
  \bibinfo{pages}{043609} (\bibinfo{year}{2001}),
  \urlprefix\url{https://link.aps.org/doi/10.1103/PhysRevA.63.043609}.

\bibitem[{\citenamefont{Sebby-Strabley
  et~al.}(2006)\citenamefont{Sebby-Strabley, Anderlini, Jessen, and
  Porto}}]{Sebby-StrabelyPRA2006}
\bibinfo{author}{\bibfnamefont{J.}~\bibnamefont{Sebby-Strabley}},
  \bibinfo{author}{\bibfnamefont{M.}~\bibnamefont{Anderlini}},
  \bibinfo{author}{\bibfnamefont{P.~S.} \bibnamefont{Jessen}},
  \bibnamefont{and} \bibinfo{author}{\bibfnamefont{J.~V.} \bibnamefont{Porto}},
  \bibinfo{journal}{Phys. Rev. A} \textbf{\bibinfo{volume}{73}},
  \bibinfo{pages}{033605} (\bibinfo{year}{2006}),
  \urlprefix\url{https://link.aps.org/doi/10.1103/PhysRevA.73.033605}.

\bibitem[{\citenamefont{F{\"o}lling et~al.}(2007)\citenamefont{F{\"o}lling,
  Trotzky, Cheinet, Feld, Saers, Widera, M{\"u}ller, and Bloch}}]{Folling2007}
\bibinfo{author}{\bibfnamefont{S.}~\bibnamefont{F{\"o}lling}},
  \bibinfo{author}{\bibfnamefont{S.}~\bibnamefont{Trotzky}},
  \bibinfo{author}{\bibfnamefont{P.}~\bibnamefont{Cheinet}},
  \bibinfo{author}{\bibfnamefont{M.}~\bibnamefont{Feld}},
  \bibinfo{author}{\bibfnamefont{R.}~\bibnamefont{Saers}},
  \bibinfo{author}{\bibfnamefont{A.}~\bibnamefont{Widera}},
  \bibinfo{author}{\bibfnamefont{T.}~\bibnamefont{M{\"u}ller}},
  \bibnamefont{and} \bibinfo{author}{\bibfnamefont{I.}~\bibnamefont{Bloch}},
  \bibinfo{journal}{Nature} \textbf{\bibinfo{volume}{448}},
  \bibinfo{pages}{1029} (\bibinfo{year}{2007}), ISSN \bibinfo{issn}{1476-4687},
  \urlprefix\url{https://doi.org/10.1038/nature06112}.

\bibitem[{\citenamefont{Trotzky et~al.}(2008)\citenamefont{Trotzky, Cheinet,
  F\"olling, Feld, Schnorrberger, Rey, Polkovnikov, Demler, Lukin, and
  Bloch}}]{TrotzkyScience2008}
\bibinfo{author}{\bibfnamefont{S.}~\bibnamefont{Trotzky}},
  \bibinfo{author}{\bibfnamefont{P.}~\bibnamefont{Cheinet}},
  \bibinfo{author}{\bibfnamefont{S.}~\bibnamefont{F\"olling}},
  \bibinfo{author}{\bibfnamefont{M.}~\bibnamefont{Feld}},
  \bibinfo{author}{\bibfnamefont{U.}~\bibnamefont{Schnorrberger}},
  \bibinfo{author}{\bibfnamefont{A.~M.} \bibnamefont{Rey}},
  \bibinfo{author}{\bibfnamefont{A.}~\bibnamefont{Polkovnikov}},
  \bibinfo{author}{\bibfnamefont{E.~A.} \bibnamefont{Demler}},
  \bibinfo{author}{\bibfnamefont{M.~D.} \bibnamefont{Lukin}}, \bibnamefont{and}
  \bibinfo{author}{\bibfnamefont{I.}~\bibnamefont{Bloch}},
  \bibinfo{journal}{Science} \textbf{\bibinfo{volume}{319}},
  \bibinfo{pages}{295} (\bibinfo{year}{2008}),
  \eprint{https://www.science.org/doi/pdf/10.1126/science.1150841},
  \urlprefix\url{https://www.science.org/doi/abs/10.1126/science.1150841}.

\bibitem[{\citenamefont{Anderlini et~al.}(2007)\citenamefont{Anderlini, Lee,
  Brown, Sebby-Strabley, Phillips, and Porto}}]{AnderliniNature2007}
\bibinfo{author}{\bibfnamefont{M.}~\bibnamefont{Anderlini}},
  \bibinfo{author}{\bibfnamefont{P.~J.} \bibnamefont{Lee}},
  \bibinfo{author}{\bibfnamefont{B.~L.} \bibnamefont{Brown}},
  \bibinfo{author}{\bibfnamefont{J.}~\bibnamefont{Sebby-Strabley}},
  \bibinfo{author}{\bibfnamefont{W.~D.} \bibnamefont{Phillips}},
  \bibnamefont{and} \bibinfo{author}{\bibfnamefont{J.~V.} \bibnamefont{Porto}},
  \bibinfo{journal}{Nature} \textbf{\bibinfo{volume}{448}},
  \bibinfo{pages}{452} (\bibinfo{year}{2007}), ISSN \bibinfo{issn}{1476-4687},
  \urlprefix\url{https://doi.org/10.1038/nature06011}.

\bibitem[{\citenamefont{Murmann et~al.}(2015)\citenamefont{Murmann,
  Bergschneider, Klinkhamer, Z\"urn, Lompe, and Jochim}}]{MurmannPRL2015}
\bibinfo{author}{\bibfnamefont{S.}~\bibnamefont{Murmann}},
  \bibinfo{author}{\bibfnamefont{A.}~\bibnamefont{Bergschneider}},
  \bibinfo{author}{\bibfnamefont{V.~M.} \bibnamefont{Klinkhamer}},
  \bibinfo{author}{\bibfnamefont{G.}~\bibnamefont{Z\"urn}},
  \bibinfo{author}{\bibfnamefont{T.}~\bibnamefont{Lompe}}, \bibnamefont{and}
  \bibinfo{author}{\bibfnamefont{S.}~\bibnamefont{Jochim}},
  \bibinfo{journal}{Phys. Rev. Lett.} \textbf{\bibinfo{volume}{114}},
  \bibinfo{pages}{080402} (\bibinfo{year}{2015}),
  \urlprefix\url{https://link.aps.org/doi/10.1103/PhysRevLett.114.080402}.

\bibitem[{\citenamefont{Dalton et~al.}(2017)\citenamefont{Dalton, Goold,
  Garraway, and Reid}}]{Dalton2017}
\bibinfo{author}{\bibfnamefont{B.~J.} \bibnamefont{Dalton}},
  \bibinfo{author}{\bibfnamefont{J.}~\bibnamefont{Goold}},
  \bibinfo{author}{\bibfnamefont{B.~M.} \bibnamefont{Garraway}},
  \bibnamefont{and} \bibinfo{author}{\bibfnamefont{M.~D.} \bibnamefont{Reid}},
  \bibinfo{journal}{Physica Scripta} \textbf{\bibinfo{volume}{92}},
  \bibinfo{pages}{023004} (\bibinfo{year}{2017}),
  \urlprefix\url{https://doi.org/10.1088/1402-4896/92/2/023004}.

\bibitem[{\citenamefont{Tichy et~al.}(2011)\citenamefont{Tichy, Mintert, and
  Buchleitner}}]{Tichy2011}
\bibinfo{author}{\bibfnamefont{M.~C.} \bibnamefont{Tichy}},
  \bibinfo{author}{\bibfnamefont{F.}~\bibnamefont{Mintert}}, \bibnamefont{and}
  \bibinfo{author}{\bibfnamefont{A.}~\bibnamefont{Buchleitner}},
  \bibinfo{journal}{Journal of Physics B: Atomic, Molecular and Optical
  Physics} \textbf{\bibinfo{volume}{44}}, \bibinfo{pages}{192001}
  (\bibinfo{year}{2011}),
  \urlprefix\url{https://doi.org/10.1088/0953-4075/44/19/192001}.

\bibitem[{\citenamefont{Benatti et~al.}(2020)\citenamefont{Benatti, Floreanini,
  Franchini, and Marzolino}}]{Benatti2020}
\bibinfo{author}{\bibfnamefont{F.}~\bibnamefont{Benatti}},
  \bibinfo{author}{\bibfnamefont{R.}~\bibnamefont{Floreanini}},
  \bibinfo{author}{\bibfnamefont{F.}~\bibnamefont{Franchini}},
  \bibnamefont{and}
  \bibinfo{author}{\bibfnamefont{U.}~\bibnamefont{Marzolino}},
  \bibinfo{journal}{Physics Reports} \textbf{\bibinfo{volume}{878}},
  \bibinfo{pages}{1} (\bibinfo{year}{2020}), ISSN \bibinfo{issn}{0370-1573},
  \bibinfo{note}{entanglement in indistinguishable particle systems},
  \urlprefix\url{https://www.sciencedirect.com/science/article/pii/S0370157320302520}.

\bibitem[{\citenamefont{Benatti et~al.}(2021)\citenamefont{Benatti, Floreanini,
  and Marzolino}}]{Benatti2021}
\bibinfo{author}{\bibfnamefont{F.}~\bibnamefont{Benatti}},
  \bibinfo{author}{\bibfnamefont{R.}~\bibnamefont{Floreanini}},
  \bibnamefont{and}
  \bibinfo{author}{\bibfnamefont{U.}~\bibnamefont{Marzolino}},
  \bibinfo{journal}{Entropy} \textbf{\bibinfo{volume}{23}}
  (\bibinfo{year}{2021}), ISSN \bibinfo{issn}{1099-4300},
  \urlprefix\url{https://www.mdpi.com/1099-4300/23/4/479}.

\bibitem[{\citenamefont{Mandel et~al.}(2003)\citenamefont{Mandel, Greiner,
  Widera, Rom, H{\"a}nsch, and Bloch}}]{MandelNature2003}
\bibinfo{author}{\bibfnamefont{O.}~\bibnamefont{Mandel}},
  \bibinfo{author}{\bibfnamefont{M.}~\bibnamefont{Greiner}},
  \bibinfo{author}{\bibfnamefont{A.}~\bibnamefont{Widera}},
  \bibinfo{author}{\bibfnamefont{T.}~\bibnamefont{Rom}},
  \bibinfo{author}{\bibfnamefont{T.~W.} \bibnamefont{H{\"a}nsch}},
  \bibnamefont{and} \bibinfo{author}{\bibfnamefont{I.}~\bibnamefont{Bloch}},
  \bibinfo{journal}{Nature} \textbf{\bibinfo{volume}{425}},
  \bibinfo{pages}{937} (\bibinfo{year}{2003}), ISSN \bibinfo{issn}{1476-4687},
  \urlprefix\url{https://doi.org/10.1038/nature02008}.

\bibitem[{\citenamefont{Daley et~al.}(2012)\citenamefont{Daley, Pichler,
  Schachenmayer, and Zoller}}]{DaleyPRL2012}
\bibinfo{author}{\bibfnamefont{A.~J.} \bibnamefont{Daley}},
  \bibinfo{author}{\bibfnamefont{H.}~\bibnamefont{Pichler}},
  \bibinfo{author}{\bibfnamefont{J.}~\bibnamefont{Schachenmayer}},
  \bibnamefont{and} \bibinfo{author}{\bibfnamefont{P.}~\bibnamefont{Zoller}},
  \bibinfo{journal}{Phys. Rev. Lett.} \textbf{\bibinfo{volume}{109}},
  \bibinfo{pages}{020505} (\bibinfo{year}{2012}),
  \urlprefix\url{https://link.aps.org/doi/10.1103/PhysRevLett.109.020505}.

\bibitem[{\citenamefont{Kaufman et~al.}(2016)\citenamefont{Kaufman, Tai, Lukin,
  Rispoli, Schittko, Preiss, and Greiner}}]{KaufmanScience2016}
\bibinfo{author}{\bibfnamefont{A.~M.} \bibnamefont{Kaufman}},
  \bibinfo{author}{\bibfnamefont{M.~E.} \bibnamefont{Tai}},
  \bibinfo{author}{\bibfnamefont{A.}~\bibnamefont{Lukin}},
  \bibinfo{author}{\bibfnamefont{M.}~\bibnamefont{Rispoli}},
  \bibinfo{author}{\bibfnamefont{R.}~\bibnamefont{Schittko}},
  \bibinfo{author}{\bibfnamefont{P.~M.} \bibnamefont{Preiss}},
  \bibnamefont{and} \bibinfo{author}{\bibfnamefont{M.}~\bibnamefont{Greiner}},
  \bibinfo{journal}{Science} \textbf{\bibinfo{volume}{353}},
  \bibinfo{pages}{794} (\bibinfo{year}{2016}), ISSN \bibinfo{issn}{0036-8075},
  \eprint{http://science.sciencemag.org/content/353/6301/794.full.pdf},
  \urlprefix\url{http://science.sciencemag.org/content/353/6301/794}.

\bibitem[{\citenamefont{Fisher et~al.}(1989)\citenamefont{Fisher, Weichman,
  Grinstein, and Fisher}}]{FisherPRB1989}
\bibinfo{author}{\bibfnamefont{M.~P.~A.} \bibnamefont{Fisher}},
  \bibinfo{author}{\bibfnamefont{P.~B.} \bibnamefont{Weichman}},
  \bibinfo{author}{\bibfnamefont{G.}~\bibnamefont{Grinstein}},
  \bibnamefont{and} \bibinfo{author}{\bibfnamefont{D.~S.}
  \bibnamefont{Fisher}}, \bibinfo{journal}{Phys. Rev. B}
  \textbf{\bibinfo{volume}{40}}, \bibinfo{pages}{546} (\bibinfo{year}{1989}),
  \urlprefix\url{https://link.aps.org/doi/10.1103/PhysRevB.40.546}.

\bibitem[{\citenamefont{Greiner et~al.}(2002)\citenamefont{Greiner, Mandel,
  Esslinger, H{\"a}nsch, and Bloch}}]{GreinerNature2002}
\bibinfo{author}{\bibfnamefont{M.}~\bibnamefont{Greiner}},
  \bibinfo{author}{\bibfnamefont{O.}~\bibnamefont{Mandel}},
  \bibinfo{author}{\bibfnamefont{T.}~\bibnamefont{Esslinger}},
  \bibinfo{author}{\bibfnamefont{T.~W.} \bibnamefont{H{\"a}nsch}},
  \bibnamefont{and} \bibinfo{author}{\bibfnamefont{I.}~\bibnamefont{Bloch}},
  \bibinfo{journal}{Nature} \textbf{\bibinfo{volume}{415}}, \bibinfo{pages}{39}
  (\bibinfo{year}{2002}), ISSN \bibinfo{issn}{1476-4687},
  \urlprefix\url{https://doi.org/10.1038/415039a}.

\bibitem[{\citenamefont{Ng et~al.}(2003)\citenamefont{Ng, Law, and
  Leung}}]{NgPRA2003}
\bibinfo{author}{\bibfnamefont{H.~T.} \bibnamefont{Ng}},
  \bibinfo{author}{\bibfnamefont{C.~K.} \bibnamefont{Law}}, \bibnamefont{and}
  \bibinfo{author}{\bibfnamefont{P.~T.} \bibnamefont{Leung}},
  \bibinfo{journal}{Phys. Rev. A} \textbf{\bibinfo{volume}{68}},
  \bibinfo{pages}{013604} (\bibinfo{year}{2003}),
  \urlprefix\url{https://link.aps.org/doi/10.1103/PhysRevA.68.013604}.

\bibitem[{\citenamefont{Ng and Leung}(2005)}]{NgPRA2005}
\bibinfo{author}{\bibfnamefont{H.~T.} \bibnamefont{Ng}} \bibnamefont{and}
  \bibinfo{author}{\bibfnamefont{P.~T.} \bibnamefont{Leung}},
  \bibinfo{journal}{Phys. Rev. A} \textbf{\bibinfo{volume}{71}},
  \bibinfo{pages}{013601} (\bibinfo{year}{2005}),
  \urlprefix\url{https://link.aps.org/doi/10.1103/PhysRevA.71.013601}.

\bibitem[{\citenamefont{Chizhov and Nazmitdinov}(2008)}]{ChizhovPRA2008}
\bibinfo{author}{\bibfnamefont{A.~V.} \bibnamefont{Chizhov}} \bibnamefont{and}
  \bibinfo{author}{\bibfnamefont{R.~G.} \bibnamefont{Nazmitdinov}},
  \bibinfo{journal}{Phys. Rev. A} \textbf{\bibinfo{volume}{78}},
  \bibinfo{pages}{064302} (\bibinfo{year}{2008}),
  \urlprefix\url{https://link.aps.org/doi/10.1103/PhysRevA.78.064302}.

\bibitem[{\citenamefont{Dutta et~al.}(2015)\citenamefont{Dutta, Barman,
  Siddharth, Khan, and Basu}}]{DuttaEPB2015}
\bibinfo{author}{\bibfnamefont{S.}~\bibnamefont{Dutta}},
  \bibinfo{author}{\bibfnamefont{A.}~\bibnamefont{Barman}},
  \bibinfo{author}{\bibfnamefont{A.}~\bibnamefont{Siddharth}},
  \bibinfo{author}{\bibfnamefont{A.}~\bibnamefont{Khan}}, \bibnamefont{and}
  \bibinfo{author}{\bibfnamefont{S.}~\bibnamefont{Basu}}, \bibinfo{journal}{The
  European Physical Journal B} \textbf{\bibinfo{volume}{88}},
  \bibinfo{pages}{139} (\bibinfo{year}{2015}), ISSN \bibinfo{issn}{1434-6036},
  \urlprefix\url{https://doi.org/10.1140/epjb/e2015-60244-9}.

\bibitem[{\citenamefont{Rubeni et~al.}(2017)\citenamefont{Rubeni, Links, Isaac,
  and Foerster}}]{RubeniPRA2017}
\bibinfo{author}{\bibfnamefont{D.}~\bibnamefont{Rubeni}},
  \bibinfo{author}{\bibfnamefont{J.}~\bibnamefont{Links}},
  \bibinfo{author}{\bibfnamefont{P.~S.} \bibnamefont{Isaac}}, \bibnamefont{and}
  \bibinfo{author}{\bibfnamefont{A.}~\bibnamefont{Foerster}},
  \bibinfo{journal}{Phys. Rev. A} \textbf{\bibinfo{volume}{95}},
  \bibinfo{pages}{043607} (\bibinfo{year}{2017}),
  \urlprefix\url{https://link.aps.org/doi/10.1103/PhysRevA.95.043607}.

\bibitem[{\citenamefont{Z\"ollner et~al.}(2008)\citenamefont{Z\"ollner, Meyer,
  and Schmelcher}}]{ZollnerPRA2008}
\bibinfo{author}{\bibfnamefont{S.}~\bibnamefont{Z\"ollner}},
  \bibinfo{author}{\bibfnamefont{H.-D.} \bibnamefont{Meyer}}, \bibnamefont{and}
  \bibinfo{author}{\bibfnamefont{P.}~\bibnamefont{Schmelcher}},
  \bibinfo{journal}{Phys. Rev. A} \textbf{\bibinfo{volume}{78}},
  \bibinfo{pages}{013621} (\bibinfo{year}{2008}),
  \urlprefix\url{https://link.aps.org/doi/10.1103/PhysRevA.78.013621}.

\bibitem[{\citenamefont{Hunn et~al.}(2013)\citenamefont{Hunn, Zimmermann,
  Hiller, and Buchleitner}}]{HunnPRA2013}
\bibinfo{author}{\bibfnamefont{S.}~\bibnamefont{Hunn}},
  \bibinfo{author}{\bibfnamefont{K.}~\bibnamefont{Zimmermann}},
  \bibinfo{author}{\bibfnamefont{M.}~\bibnamefont{Hiller}}, \bibnamefont{and}
  \bibinfo{author}{\bibfnamefont{A.}~\bibnamefont{Buchleitner}},
  \bibinfo{journal}{Phys. Rev. A} \textbf{\bibinfo{volume}{87}},
  \bibinfo{pages}{043626} (\bibinfo{year}{2013}),
  \urlprefix\url{https://link.aps.org/doi/10.1103/PhysRevA.87.043626}.

\bibitem[{\citenamefont{Dobrzyniecki and
  Sowi{\'{n}}ski}(2016)}]{DobrzynieckiEPJD2016}
\bibinfo{author}{\bibfnamefont{J.}~\bibnamefont{Dobrzyniecki}}
  \bibnamefont{and}
  \bibinfo{author}{\bibfnamefont{T.}~\bibnamefont{Sowi{\'{n}}ski}},
  \bibinfo{journal}{The European Physical Journal D}
  \textbf{\bibinfo{volume}{70}}, \bibinfo{pages}{83} (\bibinfo{year}{2016}),
  ISSN \bibinfo{issn}{1434-6079},
  \urlprefix\url{https://doi.org/10.1140/epjd/e2016-70016-x}.

\bibitem[{\citenamefont{Ishmukhamedov and
  Melezhik}(2017)}]{IshmukhamedovPRA2017}
\bibinfo{author}{\bibfnamefont{I.~S.} \bibnamefont{Ishmukhamedov}}
  \bibnamefont{and} \bibinfo{author}{\bibfnamefont{V.~S.}
  \bibnamefont{Melezhik}}, \bibinfo{journal}{Phys. Rev. A}
  \textbf{\bibinfo{volume}{95}}, \bibinfo{pages}{062701}
  (\bibinfo{year}{2017}),
  \urlprefix\url{https://link.aps.org/doi/10.1103/PhysRevA.95.062701}.

\bibitem[{\citenamefont{Zhou et~al.}(2013)\citenamefont{Zhou, Hai, Xie, and
  Tan}}]{Zhou_NJP2013}
\bibinfo{author}{\bibfnamefont{Z.}~\bibnamefont{Zhou}},
  \bibinfo{author}{\bibfnamefont{W.}~\bibnamefont{Hai}},
  \bibinfo{author}{\bibfnamefont{Q.}~\bibnamefont{Xie}}, \bibnamefont{and}
  \bibinfo{author}{\bibfnamefont{J.}~\bibnamefont{Tan}}, \bibinfo{journal}{New
  Journal of Physics} \textbf{\bibinfo{volume}{15}}, \bibinfo{pages}{123020}
  (\bibinfo{year}{2013}),
  \urlprefix\url{https://doi.org/10.1088/1367-2630/15/12/123020}.

\bibitem[{\citenamefont{Wilsmann et~al.}(2018)\citenamefont{Wilsmann, Ymai,
  Tonel, Links, and Foerster}}]{WilsmannCommPhys2018}
\bibinfo{author}{\bibfnamefont{K.~W.} \bibnamefont{Wilsmann}},
  \bibinfo{author}{\bibfnamefont{L.~H.} \bibnamefont{Ymai}},
  \bibinfo{author}{\bibfnamefont{A.~P.} \bibnamefont{Tonel}},
  \bibinfo{author}{\bibfnamefont{J.}~\bibnamefont{Links}}, \bibnamefont{and}
  \bibinfo{author}{\bibfnamefont{A.}~\bibnamefont{Foerster}},
  \bibinfo{journal}{Communications Physics} \textbf{\bibinfo{volume}{1}},
  \bibinfo{pages}{91} (\bibinfo{year}{2018}), ISSN \bibinfo{issn}{2399-3650},
  \urlprefix\url{https://doi.org/10.1038/s42005-018-0089-1}.

\bibitem[{\citenamefont{Rontani}(2013)}]{RontaniPRA2013}
\bibinfo{author}{\bibfnamefont{M.}~\bibnamefont{Rontani}},
  \bibinfo{journal}{Phys. Rev. A} \textbf{\bibinfo{volume}{88}},
  \bibinfo{pages}{043633} (\bibinfo{year}{2013}),
  \urlprefix\url{https://link.aps.org/doi/10.1103/PhysRevA.88.043633}.

\bibitem[{\citenamefont{Gharashi and Blume}(2015)}]{GharashiPRA2015}
\bibinfo{author}{\bibfnamefont{S.~E.} \bibnamefont{Gharashi}} \bibnamefont{and}
  \bibinfo{author}{\bibfnamefont{D.}~\bibnamefont{Blume}},
  \bibinfo{journal}{Phys. Rev. A} \textbf{\bibinfo{volume}{92}},
  \bibinfo{pages}{033629} (\bibinfo{year}{2015}),
  \urlprefix\url{https://link.aps.org/doi/10.1103/PhysRevA.92.033629}.

\bibitem[{\citenamefont{Sevilla et~al.}(2021)\citenamefont{Sevilla,
  Vald\'es-Hern\'andez, and Barrios}}]{SevillaQRep2021}
\bibinfo{author}{\bibfnamefont{F.~J.} \bibnamefont{Sevilla}},
  \bibinfo{author}{\bibfnamefont{A.}~\bibnamefont{Vald\'es-Hern\'andez}},
  \bibnamefont{and} \bibinfo{author}{\bibfnamefont{A.~J.}
  \bibnamefont{Barrios}}, \bibinfo{journal}{Quantum Reports}
  \textbf{\bibinfo{volume}{3}}, \bibinfo{pages}{376} (\bibinfo{year}{2021}),
  ISSN \bibinfo{issn}{2624-960X},
  \urlprefix\url{https://www.mdpi.com/2624-960X/3/3/24}.

\bibitem[{\citenamefont{Dobrzyniecki and
  Sowi\'nski}(2018)}]{DobrzynieckiPLA2018}
\bibinfo{author}{\bibfnamefont{J.}~\bibnamefont{Dobrzyniecki}}
  \bibnamefont{and}
  \bibinfo{author}{\bibfnamefont{T.}~\bibnamefont{Sowi\'nski}},
  \bibinfo{journal}{Physics Letters A} \textbf{\bibinfo{volume}{382}},
  \bibinfo{pages}{394} (\bibinfo{year}{2018}), ISSN \bibinfo{issn}{0375-9601},
  \urlprefix\url{https://www.sciencedirect.com/science/article/pii/S0375960117312070}.

\bibitem[{\citenamefont{Mal et~al.}(2019)\citenamefont{Mal, Adhikary, and
  Deb}}]{MalJPB2019}
\bibinfo{author}{\bibfnamefont{S.}~\bibnamefont{Mal}},
  \bibinfo{author}{\bibfnamefont{K.}~\bibnamefont{Adhikary}}, \bibnamefont{and}
  \bibinfo{author}{\bibfnamefont{B.}~\bibnamefont{Deb}},
  \bibinfo{journal}{Journal of Physics B: Atomic, Molecular and Optical
  Physics} \textbf{\bibinfo{volume}{52}}, \bibinfo{pages}{235001}
  (\bibinfo{year}{2019}),
  \urlprefix\url{https://doi.org/10.1088/1361-6455/ab4a72}.

\bibitem[{\citenamefont{Rungta et~al.}(2001)\citenamefont{Rungta,
  Bu\ifmmode~\check{z}\else \v{z}\fi{}ek, Caves, Hillery, and
  Milburn}}]{Rungta2001}
\bibinfo{author}{\bibfnamefont{P.}~\bibnamefont{Rungta}},
  \bibinfo{author}{\bibfnamefont{V.}~\bibnamefont{Bu\ifmmode~\check{z}\else
  \v{z}\fi{}ek}}, \bibinfo{author}{\bibfnamefont{C.~M.} \bibnamefont{Caves}},
  \bibinfo{author}{\bibfnamefont{M.}~\bibnamefont{Hillery}}, \bibnamefont{and}
  \bibinfo{author}{\bibfnamefont{G.~J.} \bibnamefont{Milburn}},
  \bibinfo{journal}{Phys. Rev. A} \textbf{\bibinfo{volume}{64}},
  \bibinfo{pages}{042315} (\bibinfo{year}{2001}),
  \urlprefix\url{https://link.aps.org/doi/10.1103/PhysRevA.64.042315}.

\bibitem[{\citenamefont{Levitin and Toffoli}(2009)}]{LevitinPRL2009}
\bibinfo{author}{\bibfnamefont{L.~B.} \bibnamefont{Levitin}} \bibnamefont{and}
  \bibinfo{author}{\bibfnamefont{T.}~\bibnamefont{Toffoli}},
  \bibinfo{journal}{Phys. Rev. Lett.} \textbf{\bibinfo{volume}{103}},
  \bibinfo{pages}{160502} (\bibinfo{year}{2009}),
  \urlprefix\url{https://link.aps.org/doi/10.1103/PhysRevLett.103.160502}.

\bibitem[{\citenamefont{Vald{\'{e}}s-Hern{\'{a}}ndez and
  Sevilla}(2020)}]{avh2020}
\bibinfo{author}{\bibfnamefont{A.}~\bibnamefont{Vald{\'{e}}s-Hern{\'{a}}ndez}}
  \bibnamefont{and} \bibinfo{author}{\bibfnamefont{F.~J.}
  \bibnamefont{Sevilla}}, \bibinfo{journal}{Journal of Physics A: Mathematical
  and Theoretical}  (\bibinfo{year}{2020}),
  \urlprefix\url{https://doi.org/10.1088/1751-8121/abcd56}.

\bibitem[{\citenamefont{Kopp}(2008)}]{KoppIJMPhy2008}
\bibinfo{author}{\bibfnamefont{J.}~\bibnamefont{Kopp}},
  \bibinfo{journal}{International Journal of Modern Physics C}
  \textbf{\bibinfo{volume}{19}}, \bibinfo{pages}{523} (\bibinfo{year}{2008}),
  \eprint{https://doi.org/10.1142/S0129183108012303},
  \urlprefix\url{https://doi.org/10.1142/S0129183108012303}.

\end{thebibliography}

\end{document}